\documentclass[12pt]{article}
\usepackage{mathrsfs}
\usepackage{natbib,graphicx,setspace}
\usepackage{mathrsfs,amsmath,amsthm,amssymb,color}
\usepackage{rotating}
\usepackage{booktabs}
\usepackage[ruled]{algorithm2e}
\usepackage{authblk}
\usepackage{bm}
\usepackage{ragged2e}
\usepackage{hyperref}
\hypersetup{
colorlinks=true,
linkcolor=blue,
citecolor=blue,
urlcolor=blue}

\usepackage{titlesec}
\titlespacing{\section}{0pt}{\parskip}{-\parskip}
\titlespacing{\subsection}{0pt}{\parskip}{-\parskip}
\titlespacing{\subsubsection}{0pt}{\parskip}{-\parskip}

\def\spacingset#1{\renewcommand{\baselinestretch}%
{#1}\small\normalsize} \spacingset{1}

\newtheorem{theorem}{Theorem}
\newtheorem{lemma}{Lemma}
\newtheorem{proposition}{Proposition}

\newtheorem{remark}{Remark}

\numberwithin{equation}{section}

\def\ve{\varepsilon}

\def\beq{\begin{equation}}
\def\eeq{\end{equation}}
\def\beqr{\begin{eqnarray}}
\def\eeqr{\end{eqnarray}}
\def\beqrs{\begin{eqnarray*}}
\def\eeqrs{\end{eqnarray*}}
\def\bet{\begin{theorem}}
\def\eet{\end{theorem}}
\def\bel{\begin{lemma}}
\def\eel{\end{lemma}}
\def\bep{\begin{proposition}}
\def\eep{\end{proposition}}
\def\bg{\begin{figure}[tbph]\begin{center}}
\def\eg{\end{center}\end{figure}}

\def\bc{\begin{center}}
\def\ec{\end{center}}

\def\wt{\widetilde}

\def\wh{\widehat}

\def\ol{\overline }

\def\mR{\mathbb{R}}

\def\mL{\mathcal L}
\def\mS{\mathbb S}

\def\mM{\mathcal M}

\def\mS{\mathcal S}

\def\cov{\mbox{cov}}

\def\dbic{\mbox{DBIC}}

\def\rmse{\mbox{RMSE}}

\def\zero{\mathbf{0}}
\def\defeq{\stackrel{\mathrm{def}}{=}}  
\def\sign{\mbox{sign}}

\begin{document}

\title{\bf Least Squares Approximation for a Distributed System}

\author[1]{Xuening Zhu}
\author[2]{Feng Li}
\author[3]{Hansheng Wang}
\affil[1]{School of Data Science, Fudan University, Shanghai, China.}
\affil[2]{School of Statistics and Mathematics, Central University of Finance and Economics, Beijing, China. }
\affil[3]{Guanghua School of Management, Peking University, Beijing, China. }

\date{}
 \maketitle
 \begin{footnotetext}[1]{Feng Li is the corresponding author. Xuening Zhu is supported by
     the National Natural Science Foundation of China (nos. 11901105, 71991472, U1811461),
     the Shanghai Sailing Program for Youth Science and Technology Excellence
     (19YF1402700), and funds from Fudan University. Feng Li's research is supported by
     the National Natural Science Foundation of China (no. 11501587). Hansheng Wang's
     research is partially supported by National Natural Science Foundation of China
     (No. 11831008). }
\end{footnotetext}

\begin{abstract}

  In this work, we develop a distributed least squares approximation (DLSA) method that is able to solve a large family of regression problems (e.g., linear regression, logistic regression, and Cox's model) on a distributed system. By approximating the local objective function using a local quadratic form, we are able to obtain a combined estimator by taking a weighted average of local estimators. The resulting estimator is proved to be statistically as efficient as the global estimator. Moreover, it requires only one round of communication. We further conduct a shrinkage estimation based on the DLSA estimation using an adaptive Lasso approach. The solution can be easily obtained by using the LARS algorithm on the master node. It is theoretically shown that the resulting estimator possesses the oracle property and is selection consistent by using a newly designed distributed Bayesian information criterion (DBIC).  The finite sample performance and computational efficiency are further illustrated by an extensive numerical study and an airline dataset. The airline dataset is 52 GB in size. The entire methodology has been implemented in Python for a {\it de-facto} standard Spark system. The proposed DLSA algorithm on the Spark system takes 26 minutes to obtain a logistic regression estimator, which is more efficient and memory friendly than conventional methods.

  \noindent {\bf KEY WORDS: } Distributed System; Least Squares Approximation; Shrinkage
  Estimation; Distributed BIC.\\

\end{abstract}

\newpage

\spacingset{1.5}
\section{Introduction}

Modern data analysis often needs to address huge datasets. In many cases, the size of the
dataset could be too large to be conveniently handled by a single computer. Consequently,
the dataset must be stored and processed on many connected computer nodes, which
thereafter are referred to as a distributed system. More precisely, a distributed system
refers to a large {cluster} of computers, which are typically connected with each other
via wire protocols such as RPC and HTTP \citep{zaharia2012resilient}. Consequently, they
are able to communicate with each other and accomplish the intended data analysis tasks at
huge scales in a collective manner.

By using a distributed system, we are able to break a large-scale computation problem into
many small pieces and then solve them in a distributed manner. A key challenge faced by
statistical computation on a distributed system is the communication cost. The
communication cost refers to the wall-clock time cost needed for data communication
between different computer nodes, which could be expensive in distributed systems
\citep{zhang2013communication,shamir2014communication,jordan2018communication}. In this
work, we consider a ``master-and-worker''-type distributed system with strong workers. We
assume that the workers are strong in the sense they are modern computers with reasonable
storage and computing capacities. For example, a worker with 32 CPU cores, 128 GB RAM, and
512 GB SSD hard disk could be a very strong worker. As one will see later, the most widely
used distributed environment, Hadoop \citep[version 2.7.2]{hadoop} and Spark
\citep[version 2.3.1]{spark}, belong to this category. Typically, the workers do not
communicate with each other directly. However, they should be connected to a common master
node, which is another computer with outstanding capacities. Consequently, most data
should be distributed to workers, and most computations should be conducted by the
workers. {This enables us to solve a large-scale computation problem in a distributed
  manner.} In contrast, the master should take the responsibility to coordinate with
different workers.

For this ``master-and-worker''-type distributed system, the communication cost is mostly
between the master and workers. One can easily verify that good algorithms for some simple
moment estimates (e.g., sample mean) can be easily developed using this type of
distributed system. For example, to compute the sample mean on a distributed system, one
can first compute the sample mean on each worker, which is known as a {map} process. Then,
each worker reports to the master the resulting sample mean and the associated sample
size. Thereafter, the master can compute the overall sample mean by a weighted average of
the sample means from each worker, which is known as a {reduce} process. Such a
``MapReduce'' algorithm requires only one ``master-and-worker'' communication for each
worker. It requires no direct communication between workers. Because most computations are
accomplished by the workers, it also makes good use of the strong worker capacities. As a
result, the algorithm can be considered effective. Unfortunately, cases such as the sample
mean are rather rare in statistical analysis. Most statistical algorithms do not have an
analytical solution (e.g., the maximum likelihood estimation of a logistic regression
model) and thus require multiple iterations (e.g., Newton-Raphson iteration or
stochastic-gradient-descent-type algorithms). These iterations unfortunately lead to
substantial ``master-and-worker'' communication, which is communicationally
expensive. Therefore, developing algorithms that are highly efficient computationally,
communicationally and statistically for distributed systems has become a problem of great
interest.



In the literature, the common wisdom for addressing a distributed statistical problem can
be classified into two categories. The first category is the ``one-shot'' (OS) or
``embarrassingly parallel'' approach, which requires only one round of
communication. Specifically, the local worker computes the estimators in parallel and then
communicate to the master to obtain an average global estimator
\citep{zhang2013communication,liu2014distributed,lee2015communication,battey2015distributed,fan2017distributed,chang2017distributed,chang2017divide}.
Although this approach is highly efficient in terms of communication, it might not achieve
the best efficiency in statistical estimation in most occasions
\citep{shamir2014communication,jordan2018communication}. The second approach includes
iterative algorithms, which require multiple rounds of communication between the master
and the workers. This approach typically requires multiple iterations to be taken so that
the estimation efficiency can be refined to match the global (or centralized) estimator
\citep{shamir2014communication,wang2017efficient,pmlr-v65-wang17a,jordan2018communication}.
In addition, see
\cite{yang2016feature,heinze2016dual,smith2018cocoa,li2019distributedFeatures} for
distributed statistical modelling methods when the data are distributed according to
features rather than samples.


The aforementioned two approaches are also studied for the sparse learning problem using
$\ell_1$ shrinkage estimation. For the first approach, \cite{lee2015communication}
investigated the distributed high-dimensional sparse regression using the OS approach by
combining local debiased $\ell_1$ estimates. \cite{battey2015distributed} revisited the
same problem but further considered distributed testing and estimation methods in a
unified likelihood framework, in which a refitted estimation is used to obtain an oracle
convergence rate.  For the second approach, both \cite{wang2017efficient} and
\cite{jordan2018communication} have developed iterative algorithms to solve the sparse
estimation problem, and they theoretically proved that the error bounds match the
centralized estimator.  Beyond the $\ell_1$ shrinkage estimation, \cite{chen2014split}
studied a penalized likelihood estimator with more general penalty function forms in a
high-dimensional setting.  However, to the best of our knowledge, there are no guarantees
that simultaneously ensure the model selection consistency \citep{fan2001variable} and
establish a criterion for consistent tuning parameter selection \citep{wang2007tuning}.
In addition, all of the above methods assume independent and identical samples stored by
each worker, which is questionable in practice because the distributed dataset might
experience great heterogeneity from worker to worker. We would like to remark that the
heterogeneity cannot be avoided because it is mainly due to the practical need to record
data across time or space (for example).
Ignoring the heterogeneity will lead to suboptimal or even biased estimation result.

In this work, we aim to develop a novel methodology to address a sparse estimation problem
with low dimensions ($p<n$, where $n$ is the local sample size).
 Under the high dimensional setting, we refer to the recent work of \cite{li2020distributed} for a distributed pre-feature screening procedure,
which consumes a communication cost of $O(p)$.
Hence, the feature dimension can be greatly reduced and then we could launch our
estimation procedure.
Specifically, we assume the data possessed by different workers are allowed to be
heterogeneous but share the same regression relationship.  The proposed method borrows the
idea of the least squares approximation \citep[LSA, ][]{wang2007unified} and can be used
to handle a large class of parametric regression models on a distributed system.
Specifically, let $Y\in\mR$ be the response of interest, let $X$ be the associated
predictor, and let $\theta \in \mR^p$ be the corresponding
regression coefficient.  The objective is to estimate the regression parameter $\theta$
and conduct a variable selection on a distributed system that has one master and many
strong workers. Assume data, denoted by $(Y_i, X_i)$, with $1\le i\le N$, that are
distributed across different workers. Further, assume that the sample size on each worker
is sufficiently large and of the same order. Under this setting, we propose a distributed
LSA (DLSA) method. The key idea is as follows:

\begin{itemize}
\item [{\sc (1)}] First, we estimate the parameter $\theta$ on each worker separately by
  using local data on distributed workers. This can be done efficiently by using standard
  statistical estimation methods (e.g., maximum likelihood estimation). By assuming that
  the sample size of each worker is sufficiently large, the resulting estimator and its
  asymptotic covariance estimate should be consistent but not statistically efficient, as
  compared with the global estimates.

\item [{\sc (2)}] Next, each worker passes the local estimator of $\theta$ and its
  asymptotic covariance estimate to the master. Because we do not consider a
  high-dimensional model setting, the communication cost in this regard should be
  negligible.

\item [{\sc (3)}] Once the master receives all the local estimators from the workers, a
  weighted least squares-type objective function can be constructed. This can be viewed as
  a local quadratic approximation of the global log-likelihood functions. As one can
  expect, the resulting estimator shares the same asymptotic covariance with the full-size
  MLE method (i.e., the global estimator) under appropriate regularity conditions.
\end{itemize}
\begin{figure}
  \centering
  \includegraphics[width=0.8\textwidth]{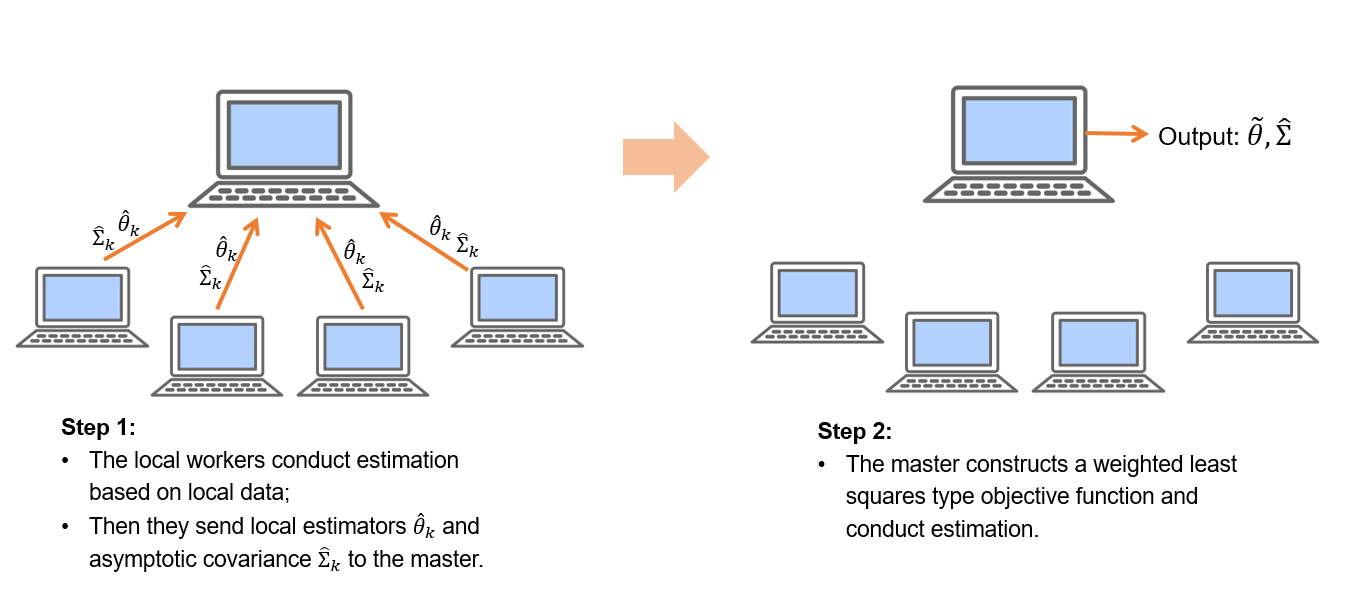}\\
\caption{\small Illustration of the DLSA method.
}\label{dlsa_steps}
\end{figure}
The major steps of the DLSA method are further illustrated in Figure \ref{dlsa_steps}.  As
one can see, the DLSA method reduces communication costs mainly by using only one round of
communication and avoids further iterative steps.  Given the DLSA objective function on
the master node, we can further conduct shrinkage estimation on the master. This is done
by formulating an adaptive Lasso-type \citep{zou2006adaptive,zhang2007adaptive} objective
function. The objective functions can be easily solved by the LARS algorithm
\citep{efron2004least} with minimal computation cost on the master. Thus, no communication
is required. Accordingly, a solution path can be obtained on the master node.
Thereafter, the best estimator can be selected from the solution path in conjunction with
the proposed distributed Bayesian information criterion (DBIC). We theoretically show that
the resulting estimation is selection consistent and as efficient as the oracle estimator,
which is the global estimator obtained under the true model.

To summarize, we aim to make the following important contributions to the existing
literature. First, we propose a master with strong workers (MSW) distributed system
framework, which solves a large-scale computation problem in a communication-efficient
way. Second, given this MSW system, we propose a novel DLSA method, which easily handles a
large class of classical regression models such as linear regression, generalized linear
regression, and {Cox's model}.
Third, due to the simple quadratic form of the objective function, the analytical solution
{path} can be readily obtained {using the LARS algorithm on the master. Then, the best
  model can be easily selected by the DBIC criterion}. Finally, but also most importantly,
the proposed DLSA method fully takes advantage of the specialty of the MSW system, which
pushes the intensive computation to the workers and therefore is as computationally,
communicationally and statistically efficient as possible.  Furthermore, we would like to
make a remark here that although the proposed DLSA is designed for a distributed system,
it can also be applied to a single computer when there are memory constraints
\citep{chen2018quantile,wang2018optimal}.

The remainder of this article is organized as follows. Section 2 introduces the model
setting and the least squares approximation method. Section 3 presents a
communication-efficient shrinkage estimation and a distributed BIC criterion. Numerical
studies are given in Section 4. An application to U.S. Airline data with datasets greater
than 52 GB is illustrated using the DLSA method on the Spark system in Section 5. The
article concludes with a brief discussion in Section 6. All technical details are
delegated to the Appendix.

\section{Statistical modelling on distributed systems}

\subsection{Model and Notations}

Suppose in the distributed system that there are in total $N$ observations, which are
indexed as $i = 1,\cdots, N$. The $i$th observation is denoted as
$Z_i = (X_i^\top, Y_i)^\top\in\mR^{p+1}$, where $Y_i\in\mR$ is the response of interest
and $X_i\in\mR^p$ is the corresponding covariate vector. Specifically, the observations
are distributed across $K$ local workers. Define $\mS = \{1,\cdots, N\}$ to be all sample
observations. Decompose $\mS = \cup_{k = 1}^K\mS_k$, where $\mS_k$ collects the
observations distributed to the $k$th worker. Obviously, we should have
$\mS_{k_1}\cap \mS_{k_2} = \emptyset$ for any $k_1\ne k_2$. Define $n=N/K$ as the average
sample size for each worker. Then, we assume $|\mS_k| = n_k$ and that all $n_k$ diverge in
the same order $O(n)$. Specifically, $c_1\le \min_k n_k/n \le \max_k n_k/n \le c_2$ for
some positive constants $c_1$ and $c_2$. We know immediately that $N = \sum n_k$. In
practice, due to the data storing strategy, the data in different workers could be quite
heterogeneous, e.g., they might be collected according to spatial regions. Despite the
heterogeneity here, we assume they share the same regression relationship, and the
parameter of interest is given by $\theta_0\in\mR^p$. We focus on the case in which $p$ is
fixed.

Let $\mL(\theta; Z)$ be a plausible twice-differentiable loss function. Define the global
loss function as $\mL(\theta) = N^{-1}\sum_{i = 1}^N\mL(\theta; Z_i)$, whose global
minimizer is $\wh\theta = \arg\min \mL(\theta)$ {and the true value is $\theta_0$.} It is
assumed that $\wh\theta$ admits the following asymptotic rule \beq
\sqrt{N}(\wh\theta-\theta_0)\rightarrow_d N(0,\Sigma)\nonumber \eeq for some positive
definite matrix $\Sigma\in\mR^{p\times p}$ {as $N\rightarrow\infty$}. If $\mL(\theta; Z)$
is the negative log-likelihood function, then $\wh \theta$ is the {global MLE estimator}.
Correspondingly, define the local loss function in the $k$th worker as
$\mL_k(\theta) = n_k^{-1}\sum_{i\in\mS_k} \mL(\theta;Z_{i})$, whose minimizer is
$\wh \theta_k = \arg\min_\theta \mL_k(\theta)$. {We assume that \beq
  \sqrt{n_k}(\wh\theta_k - \theta_0)\rightarrow_d N(0,\Sigma_k)\nonumber \eeq as
  $n_k\rightarrow\infty$ for a positive definite matrix $\Sigma_k$.} The goal is to
conduct statistical analysis based on the data on the local worker and minimize the
communication cost as much as possible.

\subsection{Least Squares Approximation and Variance Optimality}

In this section, we motivate our approach through least squares approximation to the
global loss function, which takes a local quadratic form.
To motivate this idea, we begin by decomposing and approximating the global loss function
using Taylor's expansion techniques as follows:
\begin{align}
\mL(\theta) &= N^{-1}\sum_{k = 1}^K\sum_{i\in\mS_k} \mL(\theta;Z_i)
=
N^{-1}\sum_{k = 1}^K\sum_{i\in\mS_k}
\Big\{\mL(\theta;Z_i) - \mL(\wh \theta_k;Z_i)\Big\}+C_1\nonumber\\
& \approx
N^{-1}\sum_{k = 1}^K\sum_{i\in\mS_k} (\theta - \wh \theta_k)^\top
\ddot \mL(\wh\theta_k; Z_i)(\theta - \wh\theta_k) + C_2,\label{loss_func_approx}
\end{align}
where the last equation uses the fact that $\dot \mL_k(\wh\theta_k) = 0$, and $C_1$ and
$C_2$ are some constants. Typically, the minimizer $\wh \theta_k$ will achieve the
convergence rate $\sqrt{n_k}$. Intuitively, the quadratic form in (\ref{loss_func_approx})
should be a good local approximation of the global loss function \citep{wang2007unified}.  This inspires us to
consider the following weighted least squares objective function:
\begin{align*}
\wt\mL(\theta) &= N^{-1}\sum_k(\theta - \wh\theta_k)^\top \Big\{\sum_{i\in\mS_k}\ddot \mL(\wh \theta_k; Z_i)\Big\}(\theta - \wh\theta_k),\\
&\defeq \sum_k(\theta - \wh\theta_k)^\top \alpha_k\wh\Sigma_k^{-1}(\theta - \wh \theta_k),
\end{align*}
where $\alpha_k = n_k/N$. This leads to a weighted least squares estimator (WLSE), which
takes an analytical form as follows: \beq \wt \theta = \arg\min_{\theta} \wt\mL(\theta) =
\Big(\sum_k \alpha_k\wh \Sigma_k^{-1}\Big)^{-1}\Big( \sum_k\alpha_k
\wh\Sigma_k^{-1}\wh\theta_k\Big).\label{theta_tilde} \eeq It is remarkable that the
estimator $\tilde \theta$ in (\ref{theta_tilde}) can be easily computed on a distributed
system. Specifically, the local worker sends $\wh\theta_k$ and $\wh\Sigma_k$ to the master
node, and then, the master node produces the WLSE by (\ref{theta_tilde}). As a result, the
above WLSE requires only one round of communication. Hence, it is highly efficient in
terms of communication.

Note that instead of taking a simple average of local estimators $\wh\theta_k$ in the
literature, the analytical solution in (\ref{theta_tilde}) takes a weighted average of
$\wh\theta_k$ using weights $\wh \Sigma_k^{-1}$. This will result in a higher statistical
efficiency if the data are stored heterogeneously.
To investigate the asymptotic properties of the WLSE, we assume the following conditions.

{
\begin{itemize}
\item [(C1)] {\sc (Parameter Space)} The parameter space $\Theta$ is a compact and convex
subset of $\mR^p$. In addition, the true value $\theta_0$ lies in the interior of
$\Theta$.
\item [(C2)] {\sc (Covariates Distribution)} Assume the covariates $X_i$ ($i\in \mS_k$)
from the $k$th worker are independently and identically distributed from the
distribution $F_k(x)$.
\item [(C3)] {\sc (Identifiability)} For any $\delta >0$, there exists $\epsilon > 0$
such that
\begin{align*}
  &\lim_{n\rightarrow\infty}\inf P\Big\{
  \inf_{\|\theta^* - \theta_0\|\ge \delta,1\le k \le K} \big(
  \mL_k(\theta^*) - \mL_k(\theta_0) \big)\ge \epsilon\Big\} = 1,\\
  &~~\mbox{and}~~ E\Big\{\frac{\partial\mL_k(\theta)}{\partial \theta}\Big|_{\theta = \theta_0}\Big\} = \zero.
  \end{align*}
\item [(C4)] {{\sc (Local Convexity)} Define \beq \Omega_k(\theta) = E\Big\{\frac{\partial
\mL(\theta;Z_i)}{\partial \theta}\frac{\partial \mL(\theta;Z_i)}{\partial
\theta^{\top}}\Big| i\in\mS_k\Big\} = E\Big\{\frac{\partial^2 \mL_k(\theta;
Z_i)}{\partial \theta\partial\theta^\top}\Big| i\in\mS_k\Big\} \nonumber \eeq Assume that $\Omega_k(\theta)$ is nonsingular at the true value $\theta_0$.
In addition, let $\Sigma_k = \{\Omega_k(\theta_0)\}^{-1}$, and
$\Sigma = \{\sum_k\alpha_k\Omega(\theta_0)\}^{-1}$.}
\item [(C5)] {\sc (Smoothness)} Define
$B(\delta) = \{\theta^*\in\Theta| \|\theta^* - \theta_0\|\le \delta\}$ as a ball around
the true value $\theta_0$ with radius $\delta >0$. Assume for almost all $Z\in \mR^p$
that the loss function $\mL(\theta;Z)$ admits all third derivatives
$\partial^3 \mL(\theta;Z)/(\partial \theta_i\partial\theta_j\partial\theta_l)$ for all
$\theta \in B(\delta)$. In addition, assume that there exist functions $M_{ijl}(Z)$ and
$\delta>0$ such that \beq \Big|\frac{\partial^3}{\partial
\theta_i\partial\theta_j\partial\theta_l} \mL(\theta^*;Z)\Big|\le
M_{ijl}(Z),~~~\mbox{for all}~~ \theta^*\in B(\delta), \eeq where
$E\{M_{ijl}(Z_m)| m\in\mS_k\} < \infty$ for all $1\le i,j,l\le p$ and $1\le k\le K$.
\end{itemize}

The above conditions are standard conditions to establish the asymptotic properties for
$M$-estimators. First, Condition (C1) assumes the parameter space to be convex
\citep{jordan2018communication}. Next, Condition (C2) concerns the distribution of the
covariates $\{X_i:i\in\mS_k\}$. Specifically, there are different $F_k(x)$ for different
$1\le k\le K$. In particular, it allows for the heterogeneous distribution of covariates
across workers. We would like to remark that heterogeneity is a common phenomenon in
distributed systems, and it has been ignored in much of the
literature. 
Condition (C3) assures the identifiability of the local loss functions across all
workers. Finally, Conditions (C4) and (C5) are standard regularity conditions of the loss
functions, which require certain degrees of local convexity and smoothness of the loss
functions. These conditions are widely assumed in the literature to guarantee asymptotic
convergence of the estimators
\citep{fan2001variable,lehmann2006theory,jordan2018communication}.

Given the conditions, we can establish the asymptotic properties of WLSE in the following
Proposition \ref{prop1} and Theorem \ref{thm1}.

}

\bep\label{prop1} Assume Conditions (C1)--(C5). Then, we have \beq \sqrt{N}(\wt \theta -
\theta_0) = V(\theta_0) + B(\theta_0)
\label{bias_variance}
\eeq with $\cov\{V(\theta_0)\} = \Sigma$ and $B(\theta_0) = O_p(K/\sqrt N)$, where
$\Sigma = (\sum_{k = 1}^K \alpha_k\Sigma_k^{-1})^{-1}.$ \eep The proof of Proposition
\ref{thm1} is given in Appendix A.1. Proposition \ref{thm1} {separates
  $ \sqrt{N}(\wt \theta - \theta_0)$ into two parts, namely, the variance part and the
  bias part.} Particularly, one should note that the variance order is the same as the
global estimator $\wh\theta$, which is $O(N^{-1})$, while the bias order {is related to
  the number of local workers $K$}.
   In addition, the covariance of $V(\theta_0)$ is exactly the same with the global estimator (i.e., $\Sigma$).
  Consequently, if the local sample size is sufficiently
large, the bias should be sufficiently small, and thus, the estimation efficiency
will be the same as the global estimator. We state this result in the following theorem.

\bet\label{thm1} {\sc (Global Asymptotic Normality)} Assume conditions (C1)--(C5) and
further assume {$n/N^{1/2}\rightarrow\infty$}. Then, we have
$\sqrt{N}(\wt\theta - \theta_0)\rightarrow_d N(0,\Sigma)$, which achieves the same
asymptotic normality as the global estimator $\wh\theta$. \eet The proof of Theorem
\ref{thm1} is given in Appendix A.2. It can be concluded that we should require the local
sample size to be of order larger than $\sqrt{N}$, {which is easy to satisfy in
  practice. Otherwise, we should have $N/n^2 = K/n\rightarrow \infty$. This implies that
  the number of workers is even larger than the average local sample size $n$. This is
  obviously not the case in practice
   if we have limited computational resources.}

\subsection{Two-Step Estimation}

In the previous analysis, to guarantee a $\sqrt N$-consistency rate
of the WLSE, we require that $n_k/\sqrt N\rightarrow\infty$.
The assumption might not be suitable if we have
sufficient computational resources (e.g., a large number of workers).
As a result, it could violate the condition that $K/n\rightarrow 0$
if $K$ is much larger than $n$.
To allow the case with large $K$ and small $n$, we further propose a two-step
estimator to refine $\wt\theta$.
The details are given as follows.

In the first step, we obtain WLSE $\wt\theta$ using one round of communication.
Next, the WLSE is broadcasted from the master node to the workers.
In the second step, on the $k$th work, we perform a one-step iteration using $\wt\theta$ as the initial value
to obtain
\beq
\wh \theta_k^{(2)} = \wt\theta  - \Big(\frac{\partial^2 \mL_k(\theta)}
{\partial\theta\partial\theta^\top}\Big|_{\theta = \wt\theta}
\Big)^{-1}\frac{\partial \mL_k(\theta)}{\partial \theta}\Big|_{\theta = \wt\theta}.\label{two_stepk}
\eeq
Subsequently, the local estimator $\wh\theta_k^{(2)}$
as well as $\wh \Sigma_k^{(2)} = \{\partial^2
\mL_k(\wt\theta)/\partial \theta\partial \theta^\top\}^{-1}$
are then transmitted from the worker to the master.
Similar as the design of WLSE, we obtain a two-step WLSE (TWLSE) on the master as
\beq
\wt\theta^{(2)} = \Big(\sum_k \alpha_k \wh\Sigma_k^{(2)-1}\Big)^{-1}\Big(
\sum_k\alpha_k\wh\Sigma_k^{(2)-1}\wh\theta_k^{(2)}\Big).\label{WLSE2}
\eeq
As we will show in the theoretical analysis, TWLSE enjoys
smaller bias rate than WLSE.
It is mainly because that it takes a global estimator
as an initial value with a further one-step iteration.
Hence it is capable to borrow the power of the global estimator
and further reduce the bias from $O(n^{-1})$ to $O(n^{-2})$.
We state the result rigorously in the following Theorem.
\bet\label{thm_twlse}
Assume Condition (C1)--(C5).
Then we have $\sqrt N(\wt \theta^{(2)}-\theta_0) = V_2(\theta_0) +
B_2(\theta_0)$
with $\cov\{V_2(\theta_0)\} = \Sigma$ and $B_2(\theta_0) =
O_p(n^{-2}N^{1/2})$.
In addition, we have $\sqrt{N}(\wt \theta^{(2)} - \theta_0)\rightarrow_d N(0, \Sigma)$ if we assume $n/N^{1/4}\rightarrow\infty$.
\eet
The proof of Theorem \ref{thm_twlse} is given in Appendix A.3.
The assumption $n/N^{1/4}\rightarrow\infty$ is much milder than WLSE
and in the meanwhile it allows $K = O(N^{3/4})$.
We would like to remark that the same theoretical result can be achieved if we
use one-shot estimator in our first step to reduce the communication cost.

One could see that our TWLSE can be easily extended to $m$-WLSE with $m$ steps of
iterations.
Specifically, we could repeat step (\ref{two_stepk}) and (\ref{WLSE2}) for $m$ rounds
to obtain $\wh \theta_k^{(m)}$ and $\wt\theta^{(m)}$ respectively.
Particularly, if we let $m /\log N\rightarrow\infty$, we could allow for an extreme case
that $n = O(1)$.
This is suitable for the case when we have memory constraint of local workers.
The theoretical result is given in the following Proposition.
\bep\label{mwlse}
Assume Condition (C1)--(C5) and $m /\log N\rightarrow\infty$.
Then we have $\sqrt{N}(\wt \theta^{(m)} - \theta_0)\rightarrow_d N(0, \Sigma)$.
\eep
The proof of Proposition \ref{mwlse} is given in Appendix A.4.
To achieve the desired property, we trade off
the communication cost to reduce local computational burden with
a large number of workers.

\section{Communication-efficient shrinkage estimation}

\subsection{Distributed Adaptive Lasso Estimation and Oracle Property}

Variable selection is a classical but critically important problem. That is because in
practice, the number of available covariates is typically large, but only a small number
of covariates are related to the response. Given an appropriate variable selection
technique, one can discover the important variables with high probability. In recent
decades, various variable selection techniques have been well studied
\citep{tibshirani1996regression,fan2001variable,zou2006adaptive,wang2007unified,zhang2010nearly}.
However, how to conduct variable selection on a distributed system has not been
sufficiently investigated.
Existing approaches mostly focus on the $\ell_1$ shrinkage estimation and develop
corresponding algorithms
\citep{lee2015communication,battey2015distributed,wang2017efficient,jordan2018communication}.
However, to the best of our knowledge, there are three problems that remain unsolved on a
distributed system: (a) most works do not establish the oracle properties of the shrinkage
estimators, (b) no consistent tuning parameter selection criterion is given or
investigated, and (c) the computation will be heavy if one needs to conduct estimation and
select the tuning parameters simultaneously.

To solve the above problems, we first define some notations. Without loss of generality,
we assume the first $d_0$ ($0<d_0 <p$) to be nonzero, i.e., $\theta_j \ne 0$ for
$1\le j \le d_0$ and $\theta_j = 0$ for $j >d_0$. Correspondingly, we denote
$\mM_T = \{1,\cdots, d_0\}$ to be true model. In addition, let $\mM = \{i_1,\cdots, i_d\}$
be an arbitrary candidate model with size $|\mM| = d$. In addition, for an arbitrary
vector $v = (v_j: 1\le j\le p)^\top$, define
$v^{(\mM)} = (v_i: i\in \mM)^\top\in\mR^{|\mM|}$ and
$v^{(-\mM)} = (v_i:i\not\in \mM)^\top\in\mR^{p-|\mM|}$. For an arbitrary matrix
$M = (m_{ij})$, define $M^{(\mM)} = (m_{j_1j_2}: j_1,j_2\in\mM)\in\mR^{|\mM|\times |\mM|}$.

For simultaneous variable selection and parameter estimation, we follow the idea of
\cite{wang2007unified} and consider the adaptive Lasso objective function on the master
\citep{zou2006adaptive,zhang2007adaptive}, \beq Q_\lambda(\theta) = \wt\mL(\theta) +
\sum_j \lambda_j |\theta_j|.\label{aLasso} \eeq By the adaptive Lasso method, different
amounts of shrinkage $\lambda_j$ are imposed on each estimator to improve the estimation
efficiency \citep{zou2006adaptive,zou2008one}. Compared to the LSA approach of
\cite{wang2007unified}, we have the following key differences. First, $\wt\theta$ is the
combined WLSE from local workers. Second, $\wh \Sigma$ is constructed by the local
asymptotic covariance estimators $\wh\Sigma_k$. Consequently, to achieve a global
convergence rate, one needs to carefully balance the local convergence rate of
$\wh\theta_k$ and $\wh\Sigma_k$ with that of the global ones.

\begin{remark}
Under the high dimensional setting with large $p$, the communication of $\wh\Sigma_k$ is costly
and inefficient.
Under such a case, we recommend to conduct a pre-feature screening procedure \citep{li2020distributed} to screening important features
by using one round of communication.
That consumes a communication cost of $O(p)$.
This enables us to reduce the feature dimension $p$ greatly to a feasible size.
Next, we could communicate the covariance estimator of important features to the master to complete
shrinkage estimation using (\ref{aLasso}),
which is communication friendly under the high dimensional setting.
We give the details of the pre-feature screening procedure in Section 4.2.
\end{remark}

Define $\wt\theta_\lambda = \arg\min_\theta Q_\lambda(\theta)$. Then, we can establish
the $\sqrt{N}$-consistency as well as the selection consistency result of
$\wt\theta_\lambda$ under certain conditions of
$\lambda = (\lambda_1,\cdots,\lambda_p)^\top$.

\bet\label{consistency} Assume the conditions (C1)--(C5)
 and $n/N^{1/2}\rightarrow \infty$.
Let $a_\lambda = \max\{\lambda_j, j\le d_0\}$ and $b_\lambda = \min\{\lambda_j, j>d_0\}$. Then, the following result holds.\\
a. {\sc ($\sqrt{N}$-consistency).} If $\sqrt{N}a_\lambda\rightarrow_p 0$, then $\wt\theta_\lambda - \theta = O_p(N^{-1/2})$.\\
b. {\sc (Selection Consistency).} If $\sqrt{N}a_\lambda\rightarrow_p 0$ and
$\sqrt N b_\lambda \rightarrow_p \infty$, then \beq P(\wt \theta_{\lambda}^{(-\mM_T)} =
0)\rightarrow 1\label{sel_consis} \eeq \eet

The proof of Theorem \ref{consistency} is given in Appendix B.1. Note here that
$a_\lambda$ controls the largest amount of penalty on the true nonzero
parameters. Consequently, this amount cannot be too large; otherwise, it will result in a
highly biased estimator. In contrast, $b_\lambda$ is responsible for producing sparse
solutions of irrelevant covariates. Therefore, $b_\lambda$ should be sufficiently large to
produce an effective amount of shrinkage.

\begin{remark}
  Note that the objective function of (\ref{aLasso}) constitutes of two parts.
  The first part is the weighted least squares type objective function, and the second part is a penalty term.
  Hence the Theorem \ref{consistency} and the following theoretical properties are established based on Theorem \ref{thm1}.
  If we could allow for more steps of iterations (with TWLSE or $m$-WLSE), we could further relax the condition about $n$ here
  as in Theorem \ref{thm_twlse} and Proposition \ref{mwlse}.
\end{remark}

By Theorem \ref{consistency}, we know that with probability tending to one, we have
$\wt \theta_\lambda^{(-\mM_T)}= \zero$. Meanwhile,
$\wt\theta_\lambda^{(\mM_T) } - \theta_0^{(\mM_T)} = O_p(N^{-1/2})$. It is then natural to
ask whether the statistical efficiency of $\wt\theta_\lambda^{(\mM_T)}$ can be as good as
the oracle estimator, which is the global estimator obtained under the true model, i.e.,
$\wh\theta_{oracle} = \arg\min_{\theta \in\mR^p,\theta_j = 0, \forall j\not\in \mM_T}
\mL(\theta)$. To this end, we require the following technical condition.

\begin{itemize}
\item [(C6)] {(\sc Covariance Assumption)} Define the global unpenalized estimator as
  $\wh\theta_\mM = \arg\min_{\{\theta \in \mR^p: \theta_j = 0, \forall j\not\in \mM\}}
  \mL(\theta).$ Assume for the global estimator $\wh\theta_\mM$ with $\mM\supset \mM_T$
  that
  $\sqrt{N}(\wh\theta_\mM^{(\mM)} - \theta_0^{(\mM)})\rightarrow_d N(0, \Sigma_\mM) =
  N(0, \Omega_\mM^{-1})$, where $\Sigma_{\mM}\in\mR^{|\mM|\times |\mM|}$ is a
  positive-definite matrix,
  and $ \wh\theta_{\mM}^{(\mM)} = (\wh \theta_{\mM,j}: j\in \mM)^\top\in \mR^{|\mM|}$.
   Further assume for any $\mM \supset \mM_T$ that
  $\Omega_\mM = \Omega_{\mM_F}^{(\mM)}$ holds, where $\mM_F = \{1,\cdots, p\}$ denotes the
  whole set.
\end{itemize}

\noindent
Condition (C6) does not seem very intuitive. Nevertheless, it is a condition that is well
satisfied by most maximum likelihood estimators.
 For instance, consider a logistic regression model with response $Y_i\in \{0,1\}$, i.e.,
\begin{align*}
P(Y_i = 1|X_i)\defeq p(X_i^{(\mM_T)}) = \frac{\exp(\theta_0^{(\mM_T)\top}X_i^{(\mM_T)})}
{1+\exp(\theta_0^{(\mM_T)\top}X_i^{(\mM_T)})}.
\end{align*}
For an overfitted model, the inverse asymptotic covariance of $\sqrt N\wh\theta_\mM^{(\mM)}$ is
$\Omega_{\mM} = \sum_{k = 1}^K \alpha_k E\{$ $
p(X_i^{(\mM_T)})(1-p(X_i^{(\mM_T)}))X_i^{(\mM)}X_i^{(\mM)\top}\}$,
which is a submatrix of $\Omega_{\mM_F} =
\sum_{k = 1}^K \alpha_k E\{
p(X_i^{(\mM_T)})$ $(1-p(X_i^{(\mM_T)}))X_i^{(\mM_F)}X_i^{(\mM_F)\top}\}$.
Hence Condition (C6) holds.
By similar argument, one can show that for any regression model with likelihood function of the form $\prod_i f(Y_i, X_i^\top\theta)$, the Condition (C6) is satisfied.
A more detailed discussion has been
provided by \cite{wang2007unified}. We then have the oracle property in the following
theorem.

\bet\label{oracle}
{\sc (Oracle Property)}
Assume Conditions (C1)--(C6)
 and $n/N^{1/2}\rightarrow \infty$.
Let $\sqrt{N}a_\lambda \rightarrow_p 0$ and $\sqrt{N}b_\lambda \rightarrow_p \infty$; then, it holds that
\beq
\sqrt{N}\Big(\wt\theta_{\lambda}^{(\mM_T)} - \theta^{(\mM_T)}\Big)\rightarrow_d N\Big(0, \Sigma_{\mM_T}\Big).\label{normality}
\eeq
\eet
By Theorem \ref{consistency} and \ref{oracle}, we know that as long as the tuning
parameters are approximately selected, the resulting estimator is selection consistent and
as efficient as the oracle estimator. It is remarkable that tuning a total of $p$
parameters simultaneously is not  feasible in practice. To fix this problem, we follow
the tradition of \cite{zou2006adaptive} and \cite{wang2007tuning} to specify $\lambda_j =
\lambda_0|\wt\theta_j|^{-1}$.  Since $\wt\theta_j$ is $\sqrt N$-consistent, then as long as
as $\lambda_0$ satisfies the condition $\lambda_0 \sqrt N\rightarrow 0$ and
$\lambda_0N\rightarrow\infty$, then the conditions $\sqrt N a_\lambda \rightarrow_p 0$ and
$\sqrt Nb_\lambda \rightarrow_p \infty$ are satisfied. Thereafter, the original problem
of tuning parameter selection for $\lambda$ can be replaced by selection for $\lambda_0$.

\subsection{The Distributed Bayes Information Criterion}

Although it has been shown that asymptotically, the oracle property can be guaranteed as
long as the tuning parameters are approximately selected, it is still unclear how to
conduct variable selection in practice. That motivates us to design a BIC-type criterion
that can select the true model consistently in a completely data-driven manner
\citep{zhang2007adaptive,chen2008extended,zou2009adaptive,wang2013calibrating}.
Specifically, to consistently recover the sparsity pattern, we consider a distributed Bayesian information criterion (DBIC)-based criterion as follows:
\beq
\dbic_\lambda = (\wt \theta_\lambda - \wt\theta)^\top \wh\Sigma^{-1} (\wt\theta_\lambda - \wt\theta)+ \log N\times df_\lambda ,\label{dbic}
\eeq
where  $\wh\Sigma = (\sum_{k = 1}^K \alpha_k\wh\Sigma_k^{-1})^{-1}$ and
$df_\lambda$ is the number of nonzero elements in $\wt\theta_\lambda$.

The design of the DBIC criterion is in the spirit of the BIC criterion used in
\cite{wang2007unified}.  The difference is that the DBIC uses the WLSE estimator
$\wt\theta$ and the average of distributed covariance estimators $\wh\Sigma$ to construct
the least squares objective function.  Intuitively, if $\wt\theta$ and $\wh\Sigma$
approximate the global estimator $\wh\theta = \arg\max \mL(\theta)$ and asymptotic
covariance very well, then the DBIC criterion should be able to facilitate consistent
tuning parameter selection.  Specifically, the resulting model should be selection
consistent \citep{shao1997asymptotic}.

To formally investigate the theoretical performance of DBIC, we first define some
notations.  First, we define the set of nonzero elements of $\wh\theta_\lambda$ by
$\mM_\lambda$.  Given a tuning parameter $\lambda$, $\mM_\lambda$ could be underfitted,
correctly fitted or overfitted.  We could then have the following partition:
\begin{align*}
 \mR_- = \{&\lambda \in \mR^p: \mM_\lambda \not \supset \mM_T\},~~
 \mR_0 = \{\lambda \in \mR^p: \mM_\lambda = \mM_T\},\\
 &\mR_+ = \{\lambda \in \mR^p: \mM_\lambda \supset \mM_T, \mM_\lambda \ne \mM_T\},
 \end{align*}
 where $\mR_-$ denotes the underfitted model, and $\mR_+$ denotes an overfitted model.  We
 show in the following Theorem that the DBIC can consistently identify the true model.
 \bet\label{thm_dbic} Assume Conditions (C1)--(C6)
    and $n/N^{1/2}\rightarrow \infty$.
   Define a reference tuning parameter
 sequence $\{\lambda_N\in\mR^p\}$, where the first $d_0$ elements of $\lambda_N$ are $1/N$
 and the remaining elements are $\log N/\sqrt N$.  Then, we have \beq P\big(\inf_{\lambda\in
   \mR_-\cup \mR_+} \dbic_\lambda > \dbic_{\lambda_N}\big)\rightarrow 1.\nonumber \eeq
 \eet By Theorem \ref{consistency} and \ref{oracle}, we know that with probability tending
 to one, we should have $\mM_{\lambda_N} = \mM_T$.  Consequently, the sequence $\lambda_N$
 here plays a role as a reference sequence that leads to the true model.
 Accordingly,
 Theorem \ref{thm_dbic} implies that the optimal $\lambda$ selected by the DBIC will
 consistently identify the true model.
 This is because any $\lambda$ leading to an
 inconsistent model selection result should perform worse than $\lambda_N$ in terms of
 DBIC values.
    Namely, denote the estimated set selected by DBIC is $\wh \mM_{DBIC}$,
   then Theorem \ref{thm_dbic} implies that $\wh \mM_{DBIC} = \mM_T$ with probability tending to 1.


\section{Numerical studies}

\subsection{Simulation Models and Settings}

To demonstrate the finite sample performance of the DLSA method, we conduct a number of
simulation studies in this section.
Five classical regression models are presented, and the corresponding DLSA algorithms are
implemented. For each model, we consider two typical settings to verify the numerical
performance of the proposed method. They represent two different data storing strategies
together with competing methods. The first strategy is to distribute data in a complete
random manner. Thus, the covariates on different workers are independent and identically
distributed (i.i.d). In contrast, the second strategy allows for covariate distribution on
different workers to be heterogeneous. The estimation efficiency as well as the as the
variable selection accuracy are evaluated. Examples are given as follows.

{\sc Example 1. (Linear Regression).} We first consider one of the most popular regression
analysis tools, i.e., linear regression. In particular, we generate the continuous
response $Y_i$ by a linear relationship with the covariates $X_i$ as follows: \beq Y_i =
X_i^\top \theta_0 + \epsilon_i,\nonumber \eeq where the noise term $\ve_i$ is
independently generated using a standard normal distribution $N(0,1)$. Following
\cite{fan2001variable}, the true parameter is set as
$\theta_0 = (3, 1.5, 0, 0, 2, 0, 0, 0)^\top$.

{\sc Example 2. (Logistic Regression).} The logistic regression is a classical model that
addresses binary responses \citep{hosmer2013applied}. In this example, we generate the
response $Y_i$ independently by the Bernoulli distribution given the covariate $X_i$ as
\beq P(Y_i = 1|X_i) = \frac{\exp(X_i^\top\theta_0)}{1+\exp(X_i^\top\theta_0)}.\nonumber
\eeq We follow \cite{wang2007unified} to set the true parameter
$\theta_0 = (3, 0, 0, 1.5, 0, 0, 2, 0)^\top$.

{\sc Example 3. (Poisson Regression).} In this example, we consider the Poisson
regression, which is used to model counted responses \citep{cameron2013regression}. The
responses are generated according to the Poisson distribution as \beq P(Y|X_i,\theta_0) =
\frac{\lambda^{Y_i}}{Y_i! }\exp(-\lambda), \eeq where $\lambda = \exp(X_i^\top\theta_0)$.
The true parameter $\theta_0$ is set to $(0.8, 0, 0, 1, 0, 0,-0.4, 0,0)^\top$.

For each example, two different data storage strategies are considered.  They lead to
different covariate distributions $F_x(x)$.  Specifically, the following two settings are
investigated.

\begin{itemize}
\item {\sc Setting 1 (i.i.d Covariates).} We first consider the setting in which the data
  are distributed independently and identically across the workers. Specifically, the
  covariates $X_{ij}$ ($1\le i\le N, 1\le j\le p$) are sampled from the standard normal
  distribution $N(0,1)$.

\item {\sc Setting 2 (Heterogeneous Covariates).} Next, we look at the case whereby the
  covariates distributed across each worker are heterogeneous. This is a common case in
  practice. Specifically, on the $k$th worker, the covariates are sampled from the
  multivariate normal distribution $N(\mu_k, \Sigma_k)$, where $\mu_k = k/K$, and
  $\Sigma_k = (\sigma_{k,ij}) = (\rho_k^{|j_1-j_2|})$ with $\rho_k$ sampled from
  $U[0.2, 0.3]$.
\end{itemize}

The i.i.d setting is consistent with  the existing methods such as the one-shot estimation method.
The second heterogenous data setting is not widely considered in literature but more realistic in practice.
We evaluate the numerical performances of the proposed method and existing methods under both settings.


\subsection{Pre-Feature Screening for High Dimensional Data}

For high dimensional data, transmitting covariance estimators $\wh \Sigma_k\in \mR^{p\times p}$
could be costly.
To reduce the feature dimension,
we use a pre-feature screening procedure to screening important features
before we conduct model estimation.
As a consequence, the communication cost can be well controlled.
In a recent work of \cite{li2020distributed}, they design
a distributed feature screening method using componentwise debiasing approach.
The method first expresses specific correlation measures using a non-linear
function of several component parameters,
and then conduct distributed unbiased estimation of each component parameter.
The method is shown to have better performance than simple average of correlation measures among workers.

Specifically, we express the correlation measure between the response $Y$
and the $j$th feature
as $\omega_j = g(\nu_{j1},\cdots, \nu_{js})$,
where $g$ is pre-specified function and $\nu_{j1},\cdots,\nu_{js}$
are $s$ components.
Suppose we obtain the estimation of the $\nu_{jm}$ on the $j$th worker as
$\wh \nu_{jm}^{(k)}$.
Then by distributed estimation we can obtain $\ol \nu_{jm} = K^{-1}
\sum_{k = 1}^K \wh \nu_{jm}^{(k)}$
and this enables us to resemble
$\wh \omega_j = g(\ol \nu_{j1},\cdots, \ol \nu_{js})$
on the master.
As a result, the pre-feature screening procedure will use one round of communication with communication cost of $O(p)$.
After the screening procedure, we broadcast the indexes of screened important features to workers for further estimation.
For illustration purpose, we give the following two examples to explain that how the correlation measure is calculated.
We also include the Kendall $\tau$ rank correlation \citep{li2012robust} and
SIRS correlation \citep{zhu2011model} for comparison.
See \cite{li2020distributed} for more details.

{\bf 1. Pearson Correlation.}

The Pearson correlation is used by \cite{Fan:Lv:2008}
in the sure independence screening (SIS) procedure.
In this case we have
\begin{align*}
\omega_j = g(\nu_1,\cdots, \nu_{j5}) = \left|
\frac{E(X_{ij}Y_i) - E(X_{ij})E(Y_i)}{\sqrt{(EX_{ij}^2 -
E^2(X_{ij}))(EY_i^2 - E^2(Y_i))}}\right|,
\end{align*}
where $\nu_{j1} = E(X_{ij}Y_i)$,
$\nu_{j2} = E(X_{ij})$,
$\nu_{j3} = E(Y_i)$,
$\nu_{j4} = E(X_{ij}^2)$,
and $\nu_{j5} = E(Y_i^2)$.
Using simple moments estimation we could estimate the above five components.
For instance we have
$\wh \nu_{j1}^{(k)} = n_k^{-1}\sum_{i \in \mS_k} X_{ij}Y_i$.
As a result, the estimated $\wh \omega_j$ is exactly the same with using the whole dataset in this case.

{\bf 2. Distance Correlation.}

The distance correlation (DC) is used by \cite{li2012feature} as a model-free screening index.
In this case $\omega_j$ can be expressed as
\begin{align*}
\omega_j = g(\nu_1,\cdots, \nu_8) = \frac{\nu_{j1} + \nu_{j2}\nu_{j3} - 2\nu_{j4}}{\sqrt{(\nu_{j5} +\nu_{j2}^2 -2\nu_{j6})(\nu_{j7}
+\nu_{j3}^2-2\nu_{j8})}},
\end{align*}
where $\nu_{j1} = E\{|Y_i - Y_{i}'|\cdot |X_{ij} - X_{ij}'|\}$,
$\nu_{j2} = E\{|Y_i - Y_{i}'|\}$,
$\nu_{j3} = E\{|X_{ij} - X_{ij}'|\}$,
$\nu_{j4} = E\{E(|Y_i - Y_i'|~|Y_i)E(|X_{ij} - X_{ij}'|~ |X_{ij})\}$,
$\nu_{j5} = E\{|Y_i - Y_{i}'|^2\}$,
$\nu_{j6} = E\{E^2(|Y_i - Y_{i}'|^2|Y_i)\}$,
$\nu_{j7} = E\{|X_{ij} - X_{ij}'|^2\}$,
$\nu_{j8} = E\{E^2(|X_{ij} - X_{ij}'|~|X_{ij})\}$,
and $(Y_i', X_{ij}')$ is an independent copy of $(Y_i, X_{ij})$.
On the $k$th worker,
we could estimate $\nu_{j1}$ to $\nu_{j8}$ using local dataset.
For instance, we can estimate $\nu_{j1}$ as follows,
\begin{align*}
\wh \nu_{j1}^{(k)} = \frac{1}{n_k(n_k-1)} \sum_{i_1\ne i_2, i_1,i_2\in \mS_k}
|Y_{i1} - Y_{i2}|\cdot |X_{i_1j} - X_{i_2j}|.
\end{align*}
We refer to \cite{li2012feature} for detailed estimations for $\nu_{j2}$ to $\nu_{j8}$.

Under the high dimensional setting we set $(N, p, K) = (2000, 10^3, 5)$
and $(10^4, 10^4, 8)$ respectively.
Hence the second case is more challenging with $N = p$.
Correspondingly, the non-zero coefficients are set as $\theta_{0j} = U_1\cdot \sign(U_2)$
for $1\le j\le 15$, where $U_1\sim U[0.5, 2]$ and $U_2\sim U[-0.2, 0.8]$.  The numerical
performances are evaluated for linear regression ({\sc Example 1}) and logistic regression
({\sc Example 2}) models respectively.  After the pre-screening procedure, we select top
$p_0=$40 features with highest screening measure values ($\wh \omega_j$) as our candidate
important feature set $\wh \mM_{screen}$.  Based on $\wh \mM_{screen}$, we further conduct
post-estimation with our DLSA method.  We would like to further remark that if the
post-inference is needed, the pre-feature screening procedure may introduce post-selective
bias \citep{berk2013valid,lee2014exact,taylor2015statistical,lee2016exact}.  Such bias may
not be a major issue for estimation, but is critical for statistical inference.  Hence,
selection-adjusted statistical inference should be investigated and devised.  We leave
this problem as a future research topic.



\subsection{Performance Measurements}

In this section, we give performance measurements and summarize simulation results with
respect to the estimation efficiency as well as the variable selection accuracy. The
sample sizes are set as $N = (20, 100)\times 10^3$. Correspondingly, the number of
workers is set to $K = (10, 20)$.

For a reliable evaluation, we repeat the experiment $R = 500$ times.
Take the WLSE for example.
For the $r$th
replication, denote $\wh\theta^{(r)}$ and $\wt\theta^{(r)}$ as the global estimator and
WLSE, respectively. To measure the estimation efficiency, we calculate the root mean
square error (RMSE) for the $j$th estimator as
$\rmse_{\wt\theta,j} = \{R^{-1}\sum_r \|\wt\theta_j^{(r)} - \theta_{0j}\|^2\}^{1/2}$. The
RMSE for the global estimator $\wh\theta^{(r)}$ can be defined similarly. Then, the relative
estimation efficiency (REE) with respect to the global estimator is given by
$\mbox{REE}_j = \rmse_{\wh\theta,j}/\rmse_{\wt\theta,j}$ for $j = 1,\cdots, p$.
 Next, for each $1\le j\le p$, we could construct a 95\% confidence interval for $\theta_{0j}$
as
CI$_j^{(r)} = (\wt\theta_j^{(r)} - z_{0.975}\wt{\mbox{SE}}_j^{(r)},
\wt\theta_j^{(r)} + z_{0.975}\wt{\mbox{SE}}_j^{(r)})$,
where $\wt{\mbox{SE}}_j^{(r)}$ is the $j$th diagonal element of $\wh \Sigma = (\sum_k \alpha_k \wh \Sigma_k^{-1})^{-1}$,
and $z_\alpha$ is the $\alpha$th quantile of a standard normal distribution.
Then the coverage probability (CP) is computed as CP$_j = R^{-1}\sum_{r = 1}^R I(\theta_j\in \mbox{CI}_j^{(r)})$.

Next, based on the TWLSE, we further conduct shrinkage estimation on the master node. Let
$\wh\mM^{(r)}$ be the set of selected variables in the $r$th replication using the
DBIC. Correspondingly, $\wt\theta_{\lambda}^{(r)}$ is the shrinkage estimator. To measure
the sparse discovery accuracy, we calculate the average model size (MS) as MS =
$R^{-1}\sum_{r} |\wh\mM^{(r)}|$. Next, the percentage of the true model being correctly
identified is given by CM$ = R^{-1}\sum_rI(\wh\mM^{(r)} = \mM_T)$. In addition, to further
investigate the estimation accuracy, we calculate the REE of the shrinkage estimation with
respect to the global estimator as
REE$_{j}^{s} = \rmse_{\wh\theta,j}/\rmse_{\wt\theta_{\lambda},j}$ for $j\in \mM_T$.

Lastly, for the high dimensional setting, we further evaluate the performances
of four screening measures and post-estimation result.
Specifically, we select top 40 features with highest screening measure values ($\wh \omega_j$),
which is denoted as $\wh \mM_{screen}^{(r)}$ in the $r$th replication.
The coverage rate is defined as CR$_{screen} = R^{-1}\sum_{r = 1}^R I(\mM_T\subset \wh \mM_{screen}^{(r)})$.
Similarly, one could define CR$_{select}$ for the post shrinkage estimation result.
Due to the high dimensionality, we define an overall REE$ = (\sum_j\mbox{RMSE}_{\wh \theta, j})/(\sum_j\mbox{RMSE}_{\wt \theta, j})$
for the estimator $\wt\theta$.

\subsection{Simulation Results}

We compare the proposed DLSA method with (a) the OS estimator
\citep{zhang2013communication}, and (b) the CSL estimator
\citep{jordan2018communication}. Specifically, we denote the DLSA method with two-step estimation (TWLSE) as TDLSA method.
The simulation results for the i.i.d setting is presented in Table
\ref{tab_linear_iid}--\ref{tab_logistic_iid} (in the Appendix).
In addition, the results for the heterogenous setting
are summarized in Table
\ref{tab_linear_niid}--\ref{tab_poisson_niid}.
First, in the i.i.d case, one can observe that all
three methods are as efficient as the global estimator when $N$ is increased. For example,
for the linear regression (i.e., Table \ref{tab_linear_iid}), all the methods could achieve
$\mbox{REE} \approx 1$ when $N = 100,000$ and $K = 20$.
However, in the heterogeneous
setting (i.e., Setting 2), the finite sample performances of the competing methods
are quite different. The proposed DLSA and TDLSA method achieve higher efficiency than  the
other methods, which is also asymptotically efficient as the global estimator.
 Furthermore, the TDLSA method is more efficient than the DLSA method.
For instance, the
REEs of the TDLSA estimation for the logistic regression (i.e., Table \ref{tab_logistic_niid}) is
near 1 in the second setting with $N = 100,000$ and $K = 20$, while the REE for $\theta_1$ of the OS,
CSL, DLSA methods are approximately 0.77, 0.12, and 0.92, respectively.
Although the OS estimator is
less efficient, it is still consistent as $N$ increases.
The CSL
method  performs worst under this situation.  That is because it only uses the local Hessian
matrix; this could result in a highly biased estimator.

 Lastly, the CPs for both DLSA and TDLSA methods are all around 95\%, while the CP for the CSL method is largely underestimated
especially under the heterogenous setting.
With respect to the shrinkage estimation, one can observe that the adaptive Lasso
estimator is able to achieve higher estimation efficiency in the second setting than the
global estimator. For example, the REE for the shrinkage DLSA (SDLSA) method could be even
higher than 1 for Poisson model in Setting 2 (i.e., Table \ref{tab_poisson_niid}).
In addition, we observe that the proposed DBIC does a great job at identifying the nonzero variables with high
accuracy.

\begin{table}
  \caption{Simulation results for Example 1 (Setting 2) with 500 replications.  The numerical
    performances are evaluated for different sample sizes $N$ ($\times 10^3$) and numbers
    of workers $K$.
    The REE is reported for all estimators.
    For the CSL, DLSA, TDLSA method, the CP is further reported in parentheses.
    Finally, the
    MS and CM are reported for the SDLSA method to evaluate the variable selection
    accuracy.  }\label{tab_linear_niid}
\begin{center}
\begin{tabular}{lcccccccccc}
\toprule
 Est.  & $\theta_1$ & $\theta_2$ & $\theta_3$ & $\theta_4$ & $\theta_5$ & $\theta_6$ & $\theta_7$ & $\theta_8$ & MS   & CM   \\
  \midrule
  \multicolumn{11}{c}{$N= 20$, $K=10$}                                                                                                   \\
OS    & 0.98       & 1.00       & 0.99       & 1.00       & 1.00       & 0.99       & 0.99       & 0.98       &      &      \\
CSL   & 0.94       & 0.93       & 0.94       & 0.94       & 0.95       & 0.96       & 0.93       & 0.96       &      &      \\
      & (94.2)     & (93.6)     & (95.4)     & (92.2)     & (92.8)     & (92.6)     & (91.0)     & (91.2)     &      &      \\
DLSA  & 1.00       & 1.00       & 1.00       & 1.00       & 1.00       & 1.00       & 1.00       & 1.00       &      &      \\
      & (96.0)     & (94.8)     & (96.8)     & (94.0)     & (94.6)     & (94.0)     & (94.0)     & (93.0)     &      &      \\
TDLSA & 1.00       & 1.00       & 1.00       & 1.00       & 1.00       & 1.00       & 1.00       & 1.00       &      &      \\
      & (96.0)     & (94.8)     & (96.8)     & (94.0)     & (94.6)     & (94.0)     & (94.0)     & (93.0)     &      &      \\
SDLSA & 1.08       & -          & -          & 1.13       & -          & -          & 1.19       & -          & 3.01 & 1.00 \\
\midrule
  \multicolumn{11}{c}{$N= 100$, $K=20$}                                                                                                   \\
OS    & 1.00       & 0.99       & 1.00       & 1.00       & 1.00       & 0.99       & 1.00       & 0.99       &      &      \\
CSL   & 0.93       & 0.91       & 0.92       & 0.91       & 0.93       & 0.92       & 0.92       & 0.92       &      &      \\
      & (93.2)     & (91.0)     & (92.0)     & (92.6)     & (92.0)     & (91.6)     & (92.4)     & (90.6)     &      &      \\
DLSA  & 1.00       & 1.00       & 1.00       & 1.00       & 1.00       & 1.00       & 1.00       & 1.00       &      &      \\
      & (95.4)     & (95.4)     & (94.6)     & (94.6)     & (95.6)     & (93.8)     & (94.4)     & (94.6)     &      &      \\
TDLSA & 1.00       & 1.00       & 1.00       & 1.00       & 1.00       & 1.00       & 1.00       & 1.00       &      &      \\
      & (95.4)     & (95.4)     & (94.6)     & (94.6)     & (95.6)     & (93.8)     & (94.4)     & (94.6)     &      &      \\
SDLSA & 1.10       & -          & -          & 1.13       & -          & -          & 1.16       & -          & 3.01 & 1.00 \\
\bottomrule
\end{tabular}
\end{center}
\end{table}

\begin{table}
  \caption{Simulation results for Example 2 (Setting 2) with 500 replications.  The numerical
    performances are evaluated for different sample sizes $N$ ($\times 10^3$) and numbers
    of workers $K$.
    The REE is reported for all estimators.
    For the CSL, DLSA, TDLSA method, the CP is further reported in parentheses.
    Finally, the
    MS and CM are reported for the SDLSA method to evaluate the variable selection
    accuracy.  }\label{tab_logistic_niid}
\begin{center}
\begin{tabular}{lcccccccccc}
\toprule
 Est.  & $\theta_1$ & $\theta_2$ & $\theta_3$ & $\theta_4$ & $\theta_5$ & $\theta_6$ & $\theta_7$ & $\theta_8$ & MS   & CM   \\
  \midrule
  \multicolumn{11}{c}{$N= 20$, $K=10$}                                                                                                   \\
OS    & 0.76       & 0.87       & 0.87       & 0.79       & 0.89       & 0.91       & 0.78       & 0.88       &      &      \\
CSL   & 0.15       & 0.17       & 0.18       & 0.16       & 0.19       & 0.17       & 0.16       & 0.17       &      &      \\
      & (40.8)     & (43.0)     & (42.6)     & (42.0)     & (44.8)     & (42.0)     & (43.6)     & (40.0)     &      &      \\
DLSA  & 0.88       & 1.01       & 1.01       & 0.94       & 1.01       & 1.01       & 0.90       & 1.01       &      &      \\
      & (90.0)     & (95.8)     & (95.8)     & (92.6)     & (96.0)     & (96.0)     & (92.4)     & (95.0)     &      &      \\
TDLSA & 1.00       & 1.00       & 1.00       & 1.00       & 1.00       & 1.00       & 1.00       & 1.00       &      &      \\
      & (95.6)     & (96.0)     & (95.6)     & (95.0)     & (95.6)     & (95.4)     & (95.0)     & (95.0)     &      &      \\
SDLSA & 1.02       & -          & -          & 1.09       & -          & -          & 1.05       & -          & 3.00 & 1.00 \\
  \midrule
\multicolumn{11}{c}{$N= 100$, $K=20$}                                                                                                   \\
OS    & 0.77       & 0.95       & 0.95       & 0.84       & 0.95       & 0.93       & 0.82       & 0.93       &      &      \\
CSL   & 0.12       & 0.13       & 0.13       & 0.12       & 0.14       & 0.13       & 0.12       & 0.12       &      &      \\
      & (31.8)     & (29.0)     & (29.2)     & (29.8)     & (30.0)     & (26.2)     & (29.4)     & (29.2)     &      &      \\
DLSA  & 0.92       & 1.00       & 1.01       & 0.94       & 1.00       & 1.01       & 0.92       & 1.01       &      &      \\
      & (92.6)     & (94.6)     & (94.2)     & (93.2)     & (93.2)     & (96.0)     & (92.8)     & (95.4)     &      &      \\
TDLSA & 1.00       & 1.00       & 1.00       & 1.00       & 1.00       & 1.00       & 1.00       & 1.00       &      &      \\
      & (95.6)     & (94.8)     & (94.2)     & (94.0)     & (93.4)     & (95.6)     & (94.6)     & (95.4)     &      &      \\
SDLSA & 1.02       & -          & -          & 1.11       & -          & -          & 1.07       & -          & 3.00 & 1.00 \\
\bottomrule
\end{tabular}
\end{center}
\end{table}

\begin{table}
  \caption{Simulation results for Example 3 (Setting 2) with 500 replications.  The numerical
    performances are evaluated for different sample sizes $N$ ($\times 10^3$) and numbers
    of workers $K$.
    The REE is reported for all estimators.
    For the CSL, DLSA, TDLSA method, the CP is further reported in parentheses.
    Finally, the
    MS and CM are reported for the SDLSA method to evaluate the variable selection
    accuracy.  }\label{tab_poisson_niid}
\begin{center}
\begin{tabular}{lcccccccccc}
\toprule
  Est.  & $\theta_1$ & $\theta_2$ & $\theta_3$ & $\theta_4$ & $\theta_5$ & $\theta_6$ & $\theta_7$ & $\theta_8$ & MS   & CM   \\
  \midrule
  \multicolumn{11}{c}{$N= 20$, $K=10$}                                                                                                   \\
OS    & 0.76       & 0.87       & 0.87       & 0.79       & 0.89       & 0.91       & 0.78       & 0.88       &      &      \\
CSL   & 0.15       & 0.17       & 0.18       & 0.16       & 0.19       & 0.17       & 0.16       & 0.17       &      &      \\
      & (40.8)     & (43.0)     & (42.6)     & (42.0)     & (44.8)     & (42.0)     & (43.6)     & (40.0)     &      &      \\
DLSA  & 0.88       & 1.01       & 1.01       & 0.94       & 1.01       & 1.01       & 0.90       & 1.01       &      &      \\
      & (90.0)     & (95.8)     & (95.8)     & (92.6)     & (96.0)     & (96.0)     & (92.4)     & (95.0)     &      &      \\
TDLSA & 1.00       & 1.00       & 1.00       & 1.00       & 1.00       & 1.00       & 1.00       & 1.00       &      &      \\
      & (95.6)     & (96.0)     & (95.6)     & (95.0)     & (95.6)     & (95.4)     & (95.0)     & (95.0)     &      &      \\
SDLSA & 1.02       & -          & -          & 1.09       & -          & -          & 1.05       & -          & 3.00 & 1.00 \\
      &            &            &            &            &            &            &            &            &      &      \\  [-1.1em]
  \midrule
\multicolumn{11}{c}{$N= 100$, $K=20$}                                                                                                   \\
OS    & 0.77       & 0.95       & 0.95       & 0.84       & 0.95       & 0.93       & 0.82       & 0.93       &      &      \\
CSL   & 0.12       & 0.13       & 0.13       & 0.12       & 0.14       & 0.13       & 0.12       & 0.12       &      &      \\
      & (31.8)     & (29.0)     & (29.2)     & (29.8)     & (30.0)     & (26.2)     & (29.4)     & (29.2)     &      &      \\
DLSA  & 0.92       & 1.00       & 1.01       & 0.94       & 1.00       & 1.01       & 0.92       & 1.01       &      &      \\
      & (92.6)     & (94.6)     & (94.2)     & (93.2)     & (93.2)     & (96.0)     & (92.8)     & (95.4)     &      &      \\
TDLSA & 1.00       & 1.00       & 1.00       & 1.00       & 1.00       & 1.00       & 1.00       & 1.00       &      &      \\
      & (95.6)     & (94.8)     & (94.2)     & (94.0)     & (93.4)     & (95.6)     & (94.6)     & (95.4)     &      &      \\
SDLSA & 1.02       & -          & -          & 1.11       & -          & -          & 1.07       & -          & 3.00 & 1.00 \\
\bottomrule
\end{tabular}
\end{center}
\end{table}

Next, we summarize the simulation results for the high dimensional setting in Table \ref{screening_cp}--\ref{linear_after_scr}
for linear and logistic regression models.
Specifically, the screening accuracies of four screening methods are presented in Table \ref{screening_cp}
and the post-estimation result is given in Table \ref{linear_after_scr}.
Among all screening methods, the SIS and DC are able to achieve higher screening coverage rate.
For the post-estimation result, both TDLSA and SDLSA methods have better estimation accuracy than other methods.



\section{Application to airline data}

For illustration purposes, we study a large real-world dataset. Specifically, the dataset
considered here is the U.S. Airline Dataset. The dataset is available at
\url{http://stat-computing.org/dataexpo/2009}. It contains detailed flight information
about U.S. airlines from 1987 to 2008. The task is to predict the delayed status of a
flight given all other flight information with a logistic regression model. Each sample in
the data corresponds to one flight record, which consists of a binary response variable
for delayed status (\textsf{Delayed}), and departure time, arrival time, distance of the
flight, flight date, delay status at departures, carrier information, origin and
destination as regressors. The complete variable details are described in
Table~\ref{tab:airline}.  {The data contain six continuous variables and five categorical
  variables. The categorical variables are converted to dummies with appropriate
  dimensions. We treat the \textsf{Year} and \textsf{DayofMonth} variables as numerical to
  capture the time effects. To capture possible seasonal patterns, we also convert the
  time variables \textsf{Month} and \textsf{DayofWeek} to dummies.} Ultimately, a total of
181 variables are used in the model. The total sample size is 113.9 million
observations. This leads to the raw dataset being 12 GB on a hard drive. After the dummy
transformation described in Table~\ref{tab:airline}, the overall in-memory size is over 52
GB, even if all the dummies are stored in a sparse matrix format. Thus, this dataset can
hardly be handled by a single computer. All the numerical variables are standardized to
have a mean of zero and a variance of one.
\begin{table}%
  \caption{Variable description for the U.S. airline data. Non-categorical numerical
    variables are standardized to have mean zero and variance one. }
  \label{tab:airline}
  \centering
\begin{tabular}{lp{0.4\textwidth}p{0.35\textwidth}}
\toprule
Variable   &  Description  &  Variable used in the model     \\
\midrule
\textsf{Delayed}   & Whether the flight is delayed, 1 for Yes; 0 for No. &   Used as the response variable \\
\midrule
\textsf{Year}  & Year between 1987 and 2008           &   Used as numerical variable \\
\textsf{Month} & Which month of the year  &   Converted to $11$ dummies  \\
\textsf{DayofMonth} & Which day of the month  &   Used as numerical variable           \\
\textsf{DayofWeek} & Which day of the week  &   Converted to $6$ dummies           \\
\textsf{DepTime} & Actual departure time          &    Used as numerical variable       \\
\textsf{CRSDepTime} & Scheduled departure time & Used as numerical variable      \\
\textsf{CRSArrTime} &  Scheduled arrival time & Used as numerical variable        \\
\textsf{ElapsedTime} &Actual elapsed time & Used as numerical variable  \\
\textsf{Distance} & 	Distance between the origin and destination in miles & Used as numerical variable \\
\textsf{Carrier} & Flight carrier code for $29$ carriers & Top  $7$ carries converted to $7$ dummies  \\
\textsf{Destination} & Destination of the flight (total $348$ categories)& Top $75$ destination cities converted to $75$ dummies \\
\textsf{Origin} & Departing origin (total $343$ categories)& Top $75$ origin cities converted to $75$ dummies\\
\bottomrule

\end{tabular}
\end{table}

\subsection{The Spark System and MLE}
To demonstrate our method, we set up a Spark-on-YARN cluster on the Alibaba cloud server
(\url{https://www.alibabacloud.com/products/emapreduce}). This is a standard
industrial-level architecture setup for a distributed system.

The system consists of one master node and two worker nodes. Each node contains 32 virtual
cores, 128 GB of RAM and two 80 GB SSD local hard drives. The dataset is stored on the
Hadoop data file system (HDFS). We use such hardware specification to prevent the failure
of Spark's default algorithm due to the well-known out-of-memory issue for comparison
purposes. It is worth mentioning that our DLSA algorithm works well on workers with less
than 64GB RAM.

Because the system's RAM is larger than the raw data size, one may wonder whether the
logistic regression task can be run on a single node. Unfortunately, this is infeasible in
practice.  This is because much more memory (typically $>128$ GB with a double-precision
floating-point format) is needed for operating on matrices of such a huge size. Even for
the Spark system, a task of this magnitude cannot be directly performed using an existing
algorithm library (e.g., Spark ML) unless each worker node is equipped with at
least 128G RAM. This is because Spark is a very memory-intensive system. For example, to
compute a single distributed matrix with a size of approximately 1 GB in memory, one might
need each worker to have 2-5 GB of memory in practice. This overhead memory consumption
grows significantly as the size of the data matrix increases. For discussions, see the
Spark documentation at
\url{https://spark.apache.org/docs/latest/tuning.html#memory-tuning} and
\citet{chen2016xgboost}.

We compared our approach with Spark's builtin distributed algorithm (implemented in the
\texttt{spark.ml.classification.LogisticRegression} module).  If one insists on computing
the traditional MLE based on the entire dataset with a single node,  we employ a
stochastic gradient descent (SGD) algorithm \citep{zhang2004solving} in the Python
\texttt{scikit-learn} module \citep{scikit-learn} to allow for a memory-constraint
situation.

Fortunately, both the proposed DLSA and OS methods allow us to develop a user-friendly
Spark algorithm with minimal computer resources. Our DLSA works well for a Spark
system with only 64G RAM. As the algorithm is designed in a batch manner, it is highly
efficient under memory constraints. The algorithm is developed with the Spark Python API
(PySpark) and run on a Spark system (version $>$ 2.3) (see Algorithm~\ref{alg:spark} for
details). It can be freely downloaded from {\url{https://github.com/feng-li/dlsa}}. We
also provide the implementation of the aforementioned algorithms in Table \ref{tab:comp}
in the repository. We then use our algorithm to fit a logistic regression model for
modelling the delayed status of a flight.

\begin{algorithm}
  \caption{Spark implementation}
  \label{alg:spark}
\SetAlgoLined

\KwIn{The model function for modelling each partitioned dataset}

\KwOut{The weighted least squares estimator $\tilde{\theta}$, covariance matrix
$\wh{\Sigma}$, DBIC $\mathrm{DBIC}_{\lambda}$}
\hrule \textbf{Steps:}

\textbf{Step (1)}. Pre-determine the overall cluster available memory as $M_{ram}$, the
total number of CPU cores as $C_{cores}$, and the total data size to be processed as
$D_{total}$\;

\textbf{Step (2)} Define the number of batched chunks $N_{chunks}$ to allow for
out-of-memory data processing. We recommend that $N_{chunks}$ be at least greater than
$3\times D_{total}/M_{ram}$ in a Spark system.

\textbf{Step (3)}. Define the number of partitions
$P_{partition}= D_{total}/(N_{chunks}\times C_{cores})$.

\textbf{Step (4)}. Define a \emph{model function} whereby the input is an $n\times (p+2)$
Python Pandas DataFrame containing the response variable, covariates and partition id, and
the output is a $p\times (p+1)$ Pandas DataFrame whereby the first column is $\wh\theta_k$
and the remaining columns store $\wh{\Sigma}_k^{-1}$.

\textbf{Step (5)}.

\For{i in 1:$N_{chunks}$}{

  \textbf{(a)}. Transfer the data chunk to Spark's distributed DataFrame if the data are
  stored in another format.

  \textbf{(b)}. Randomly assign an integer partition label from $\{1,..., P_{partition}\}$
  to each row of the Spark DataFrame.

  \textbf{(c)}. Repartition the DataFrame in the distributed system if the data are not
  partitioned by the partition label.

  \textbf{(d)}. Group the Spark DataFrames by the assigned partition label.

  \textbf{(e)}. Apply the model function to each grouped dataset with Spark's
  \emph{Grouped map Pandas UDFs} API and obtain a $(pP_{partition})\times (p+1)$
  distributed Spark DataFrame $R_i$. \

}

\textbf{Step (6)}. Aggregate $R_i$ over both partitions and chunks and return the
$p\times (p+1)$ matrix $R_{final}$.

\textbf{Step (7)}. Return $\tilde{\theta}$ by Equation~\eqref{theta_tilde}, $\wh{\Sigma}$,
and $\mathrm{DBIC}_{\lambda}$ by Equation~\eqref{dbic}.

~\\

$\bullet$ Because the final step in the DLSA algorithm is carried out on the master node
and because data transformation from worker nodes to the master node is required, a
special tool called ``Apache Arrow'' (\url{https://arrow.apache.org/}) is plugged-in to
our system to allow efficient data transformation between Spark’s distributed DataFrame
and Python's Pandas DataFrame.

\end{algorithm}

\begin{table}%
  \caption{Log likelihood, computational time and communication time comparison with the
    airline data. Note that the inevitable overhead for initializing the spark task is
    also included in the computational time. It is difficult to monitor the exact
    communication time per code interaction in Spark due to its lazy evaluation technique
    used. The overall communication time is estimated from the computational time by
    subtracting the computational time used on the master machine and the average
    computational time used on the workers.}
  \label{tab:comp}
  \centering
  \resizebox{\textwidth}{!}{
  \begin{tabular}{lrp{3cm}p{3cm}p{2cm}p{1.5cm}}
    \toprule
    Algorithm      & Log likelihood     & Computational time (min) & Communication time (min) & Worker's RAM & No. of workers \\
    \midrule
    DLSA           & $-1.62\times10^8$  & 26.2                     & 3.8                      & 64GB         & 2              \\
    TDLSA          & $-1.62\times10^8$  & 28.7                     & 4.4                      & 64GB         & 2              \\
    OS             & $-1.65\times10^8$  & 25.3                     & 2.3                      & 64GB         & 2              \\
    Spark ML       & $-1.63\times10^8$  & 54.7                     & 18.6                     & 128GB        & 2              \\
    Serialized SGD & $-1.71 \times10^8$ & 913.2                    & ---                      & 64GB         & 1              \\
    \bottomrule
  \end{tabular}
}
\end{table}

To this end, the entire dataset is randomly partitioned into 1139 subgroups. The sample
size for each subgroup is approximately $100,000$. Next, for each subgroup of data, we
create a virtual worker (i.e., an executor in the Spark system) so that the computation
for each worker can be conducted in a parallel manner. By doing so, the computation power
of the entire Spark system can be maximized for both DLSA and OS methods. The
corresponding log-likelihood values, computational time as well as the estimated
communication time are reported in Table \ref{tab:comp}, respectively. We remark that the
computing time for the MLE includes the data shuffling time with our DLSA and OS
algorithms. This serves as an important benchmark to gauge the performance of the other
competing methods (e.g., DLSA and OS methods).  Comparing these results with that of the
traditional MLE, we find that the traditional MLE is extremely difficult to compute
without a distributed system.

Table \ref{tab:comp} also depicts that the log-likelihood value of the DLSA is the best
with an affordable RAM consumption. The Spark's ML module takes longer time than DLSA
because DLSA only requires one round communication and the parameter optimization is done
within each executor. A serialized SGD takes more that $15$ hours and obtains an inferior
result (i.e., smaller log-likelihood value).

\subsection{Variable Selection Results with BIC}

We next apply the proposed shrinkage DLSA method (referred to as SDLSA) with the BIC
criterion to conduct variable selection. It is remarkable that this can be fully conducted
on the master, and no further communication is needed. It takes only $0.2$ seconds to
accomplish the task. After the shrinkage estimation, we are able to reduce the 181
variables to 157 variables.

The detailed results are summarized in Table~\ref{tab:coef}. First, with respect to time
effects, both yearly and seasonal trends are found. The coefficient for the \textsf{Year}
is 6.12, which implies that as the year proceeds, the airline delays become more severe.
Next, the passengers are likely to encounter delays in May, October, November, and
December (with coefficients of 6.03, 0.8, 0.4, and 0.19, respectively). In terms of
days of the week, more delays are expected for certain working days, i.e., Tuesday,
Wednesday and Friday (with coefficients of 0.6, 0.85 and 0.39, respectively) compared to
weekends. Finally, within a given day, we find the coefficients for both the scheduled
departure time (\textsf{CRSDepTime}, $-0. 12$) and the scheduled arrival time
(\textsf{CRSArrTime}, $-0. 13$) are negative, indicating that late departing and arriving
flights suffer less so from delays.

Next, with respect to the airline carriers, AA (American Airlines), DL (Delta Air Lines),
NW (Northwest Airlines), and UA (United Airlines) have estimated coefficients of 0.49,
0.18, 0.39, and 0.70, which indicates how more likely they are to be delayed compared with
other airlines.  In addition, one may be interested in with which airports are more likely
to have delayed flights. According to our estimation results, the top five origin airports
that cause delays are IAH (George Bush Intercontinental Airport), LGA (LaGuardia Airport),
PHL (Philadelphia International Airport), RDU (Raleigh-Durham International Airport), ONT
(Ontario International Airport), and SMF (Sacramento International Airport), with
coefficients of $0. 82$, $0. 87$, $0. 94$, $1.58$, and $1.59$, respectively. The top five
destination airports that cause delays are PBI (Palm Beach International Airport), MCI
(Kansas City International Airport), DCA (Ronald Reagan Washington National Airport), SAN
(San Diego International Airport), and MEM (Memphis International Airport), with
coefficients of $0. 99$, $1. 00$, $1.07$, $1.15$, and $1.16$, respectively.

\begin{sidewaystable}
  \caption{Coefficients for the logistic model estimated with SDLSA using the BIC. The
    carrier and airport abbreviations are assigned by the International Air Transport
    Association. We denote ``Airport'', ``Origin'' and ``Destination'' by ``A'', ``O'' and
    ``D'', respectively. The variances for all coefficients are all smaller than
    $0.001$. The notation *** indicates 0.001 level of significance. }
  \label{tab:coef}
  \centering
\resizebox{\textwidth}{!}{
\begin{tabular}{lrrrrrrrrrrr}
\toprule
            & Intercept    & Year         & CRSArrTime & Distance       & CRSDepTime  & DayofMonth & ElapsedTime & DepTime                                \\
            & $-0.30$      & $6.12^{***}$ & $-0. 13$   & $-5. 68^{***}$ & $-0. 12$    & $-0. 06$   & $0. 08$     & $0. 15$                                \\
\midrule
Month       & Feb          & Mar          & Apr        & May            & Jun         & Jul        & Aug         & Sep     & Oct         & Nov   & Dec    \\
            & $-0.13$      & $-0.01$      & $0.03$     & $6.03^{***}$   & $-0.28$     & $-0.09$    & $-0.02$     & $-0.07$ & $0.8^{***}$ & $0.4$ & $0.19$ \\
\midrule
Day of Week & Tue          & Wed          & Thu        & Fri            & Sat         & Sun                                                               \\
            & $0.6^{***}$  & $0.85^{***}$ & $-0.57$    & $0.39$         & $0.22$      & $0.26$                                                            \\
\midrule
Carrier     & AA           & CO           & DL         & NW             & UA          & US         & WN                                                   \\
            & $0.49^{***}$ & $-0.16$      & $0.18$     & $0.39$         & $0.7^{***}$ & $-0.43$    & $-0.6$                                               \\
\midrule
A           & ABQ          & ANC          & ATL        & AUS            & BDL         & BHM        & BNA         & BOS     & BUF         & BUR   & BWI    \\

O & $ -0.46 $ & $ -0.14 $ & $ -0.02 $ & $ 0.39 $  & $ 0.38 $  & $ 0.13 $  & $ -0.17 $ & $ 0.13 $ & $ -0.55 $ & $ -0.01 $ & $ 0.54 $ \\
D & $ -0.61 $ & $ 0.78 $  & $ -0.84 $ & $ -0.73 $ & $ -0.74 $ & $ -0.46 $ & $ -0.28 $ & $ 0.51 $ & $ -0.50 $ & $ -0.90 $ & $ 0.69$  \\

A & CLE & CLT & CMH & CVG & DAL & DAY & DCA & DEN & DFW & DTW & ELP \\

O & $ 0.49 $ & $ -0.87 $ & $ -0.80 $ & $ 0.29 $  & $ -0.08 $ & $ 0.40 $  & $ -0.07 $      & $ -0.22 $ & $ 0.53 $  & $ 0.32 $  & $ -0.56 $ \\
D & $ 0.14 $ & $ -0.46 $ & $ 0.64 $  & $ -1.60 $ & $ -0.72 $ & $ -0.79 $ & $ 1.07^{***} $ & $ 0.37 $  & $ -1.04 $ & $ -0.19 $ & $ -0.54 $ \\

A & EWR & FLL & GSO & HNL & HOU & IAD & IAH & IND & JAX & JFK & LAS \\

O & $ 0.21 $  & $ -0.99 $ & $ 0.11 $ & $ 0.19 $ & $ -0.51 $ & $ -0.29 $ & $ 0.82^{***} $ & $ 0.46 $  & $ -0.71 $ & $ 0.07 $  & $ -0.94$  \\
D & $ -0.44 $ & $ -1.70 $ & $ 0.77 $ & $ 0.79 $ & $ -1.62 $ & $ 0.00 $  & $ -0.54 $      & $ -0.27 $ & $ -0.12 $ & $ -0.70 $ & $ -0.29 $ \\

A & LAX & LGA & MCI & MCO & MDW & MEM & MIA & MKE & MSP & MSY & OAK \\

O & $ 0.38 $  & $ 0.87^{***} $ & $ 0.61 $       & $ 0.52 $ & $ -0.48 $ & $ 0.71 $       & $ 0.39 $  & $ -0.64 $ & $ -0.10 $ & $ 0.21 $ & $ -1.04$ \\
D & $ -0.20 $ & $ 0.13 $       & $ 1.00^{***} $ & $ 0.30 $ & $ 0.10 $  & $ 1.16^{***} $ & $ -0.11 $ & $ -0.83 $ & $ -0.61 $ & $ 0.97 $ & $ -1.47$ \\

A & OKC & OMA & ONT & ORD & ORF & PBI & PDX & PHL & PHX & PIT & PVD \\

O & $ -0.71 $ & $ -0.85 $ & $ 1.58^{***} $ & $ -0.41 $ & $ -0.67 $ & $ -1.12 $      & $ -0.13 $ & $ 0.94^{***} $ & $ 0.37 $ & $ 0.18 $ & $ -0.42$ \\
D & $ 0.69 $  & $ -0.74 $ & $ -1.49 $      & $ -0.72 $ & $ -0.79 $ & $ 0.99^{***} $ & $ -0.64 $ & $ -0.54 $      & $ 0.83 $ & $ 0.33 $ & $ -0.40$ \\

A & RDU & RIC & RNO & ROC & RSW & SAN & SAT & SDF & SEA & SFO & SJC \\

O & $ -5.80 $ & $ 0.02 $ & $ -0.06 $ & $ 0.20 $  & $ 0.05 $ & $ 0.00 $       & $ -0.04 $ & $ 0.00 $  & $ 0.24 $  & $ 0.24 $  & $ 0.14$ \\
D & $ 0.97 $  & $ 0.83 $ & $ 0.71 $  & $ -1.07 $ & $ 0.37 $ & $ 1.15^{***} $ & $ 0.78 $  & $ -0.35 $ & $ -0.79 $ & $ -0.51 $ & $ 0.21$ \\

A & SJU & SLC & SMF & SNA & STL & SYR & TPA & TUL & TUS &  & \\

O & $ -0.14 $ & $ 0.59 $ & $ 1.59^{***} $ & $ -0.15 $ & $ 0.14 $ & $ -1.21 $ & $ -0.28 $ & $ -0.36 $ & $ -1.51 $ &  & \\
D & $ 0.84 $  & $ 0.09 $ & $ -0.03 $      & $ 0.02 $  & $ 0.38 $ & $ -0.06 $ & $ 0.31 $  & $ 0.43 $  & $ 0.03 $  &  & \\

\bottomrule
\end{tabular}
}
\end{sidewaystable}

\section{Concluding remarks}

In this article, we develop a novel DLSA algorithm that is able to perform large-scale
statistical estimation and inference on a distributed system.  The DLSA method can be
applied to a large family of regression models (e.g., logistic regression, Poisson
regression, and Cox's model).  First, it is shown that the DLSA estimator is as
statistically optimal as the global estimator.  Moreover, it is computationally efficient
and only requires one round of communication.

Furthermore, we develop the corresponding shrinkage estimation by using an adaptive Lasso
approach.  The oracle property is theoretically proven.  A new DBIC measure for
distributed variable selection, which only needs to be performed on the master and
requires no further communication, is designed.  We prove the DBIC measure to be selection
consistent.  Finally, numerical studies are conducted with five classical regression
examples.  In addition, a Spark toolbox is developed, which is shown to be computationally
efficient both through simulation and in airline data analysis.

To facilitate future research, we now discuss several interesting topics.  First, the DLSA
method requires the objective function to have continuous second-order derivatives.  This
assumption might be restrictive and cannot be satisfied for certain regression models,
e.g., the quantile regression.  Consequently, the relaxation of this assumption can be
investigated, and corresponding distributed algorithms should be designed for such
regression models.  Second, the dimension considered in our framework is finite.  As a
natural extension, one could study the shrinkage estimation properties in high-dimensional
settings. Furthermore, as we commented on our pre-feature screening procedure, further
post-selection statistical inference should be investigated
\citep{berk2013valid,taylor2015statistical,lee2016exact}.  Third, the algorithm is
designed for independent data.  In practice, dependent data (e.g., time series data and
network data) are frequently encountered.  It is thus interesting to develop corresponding
algorithms by considering the dependency structure.  Lastly, note that the penalty term
given in (\ref{aLasso}) can be added to other surrogate loss functions as in
\cite{jordan2018communication}.  Hence the theoretical analysis in Section 3 as a great
potential to extend to other methodologies.

\section{Acknowledgements}

The authors thank the editor, associate editor, and two
referees for their insightful comments that have led to significant improvement of this article.
We also thank Dr. Zhihui Jin for his constructive suggestions.

\bibliographystyle{asa}
\bibliography{../references}

\begin{thebibliography}{53}
\newcommand{\enquote}[1]{``#1''}
\expandafter\ifx\csname natexlab\endcsname\relax\def\natexlab#1{#1}\fi

\bibitem[{{Apache Software Foundation}(2019{\natexlab{a}})}]{hadoop}
{Apache Software Foundation} (2019{\natexlab{a}}), \enquote{Apache Hadoop
  (version 2.7.2),} .

\bibitem[{{Apache Software Foundation}(2019{\natexlab{b}})}]{spark}
--- (2019{\natexlab{b}}), \enquote{Apache Spark (version 2.3.1),} .

\bibitem[{Battey et~al.(2015)Battey, Fan, Liu, Lu, and
  Zhu}]{battey2015distributed}
Battey, H., Fan, J., Liu, H., Lu, J., and Zhu, Z. (2015), \enquote{Distributed
  estimation and inference with statistical guarantees,} \textit{arXiv preprint
  arXiv:1509.05457}.

\bibitem[{Berk et~al.(2013)Berk, Brown, Buja, Zhang, Zhao,
  et~al.}]{berk2013valid}
Berk, R., Brown, L., Buja, A., Zhang, K., Zhao, L., et~al. (2013),
  \enquote{Valid post-selection inference,} \textit{The Annals of Statistics},
  41, 802--837.

\bibitem[{Cameron and Trivedi(2013)}]{cameron2013regression}
Cameron, A.~C. and Trivedi, P.~K. (2013), \textit{Regression analysis of count
  data}, vol.~53, Cambridge university press.

\bibitem[{Chang et~al.(2017{\natexlab{a}})Chang, Lin, and
  Wang}]{chang2017divide}
Chang, X., Lin, S.-B., and Wang, Y. (2017{\natexlab{a}}), \enquote{Divide and
  conquer local average regression,} \textit{Electronic Journal of Statistics},
  11, 1326--1350.

\bibitem[{Chang et~al.(2017{\natexlab{b}})Chang, Lin, and
  Zhou}]{chang2017distributed}
Chang, X., Lin, S.-B., and Zhou, D.-X. (2017{\natexlab{b}}),
  \enquote{Distributed semi-supervised learning with kernel ridge regression,}
  \textit{The Journal of Machine Learning Research}, 18, 1493--1514.

\bibitem[{Chen and Chen(2008)}]{chen2008extended}
Chen, J. and Chen, Z. (2008), \enquote{Extended Bayesian information criteria
  for model selection with large model spaces,} \textit{Biometrika}, 95,
  759--771.

\bibitem[{Chen and Guestrin(2016)}]{chen2016xgboost}
Chen, T. and Guestrin, C. (2016), \enquote{Xgboost: A scalable tree boosting
  system,} in \textit{Proceedings of the 22nd acm sigkdd international
  conference on knowledge discovery and data mining}, ACM, pp. 785--794.

\bibitem[{Chen et~al.(2018)Chen, Liu, and Zhang}]{chen2018quantile}
Chen, X., Liu, W., and Zhang, Y. (2018), \enquote{Quantile regression under
  memory constraint,} \textit{arXiv preprint arXiv:1810.08264}.

\bibitem[{Chen and Xie(2014)}]{chen2014split}
Chen, X. and Xie, M.-g. (2014), \enquote{A split-and-conquer approach for
  analysis of extraordinarily large data,} \textit{Statistica Sinica},
  1655--1684.

\bibitem[{Efron et~al.(2004)Efron, Hastie, Johnstone, and
  Tibshirani}]{efron2004least}
Efron, B., Hastie, T., Johnstone, I., and Tibshirani, R. (2004), \enquote{Least
  angle regression,} \textit{Annals of Statistics}, 32, 407--499.

\bibitem[{Fan and Li(2001)}]{fan2001variable}
Fan, J. and Li, R. (2001), \enquote{Variable selection via nonconcave penalized
  likelihood and its oracle properties,} \textit{Journal of the American
  Statistical Association}, 96, 1348--1360.

\bibitem[{Fan and Lv(2008)}]{Fan:Lv:2008}
Fan, J. and Lv, J. (2008), \enquote{Sure independence screening for ultra-high
  dimensional feature space (with discussion),} \textit{Journal of the Royal
  Statistical Society, Series B}, 70, 849--911.

\bibitem[{Fan et~al.(2017)Fan, Wang, Wang, and Zhu}]{fan2017distributed}
Fan, J., Wang, D., Wang, K., and Zhu, Z. (2017), \enquote{Distributed
  estimation of principal eigenspaces,} \textit{arXiv preprint
  arXiv:1702.06488}.

\bibitem[{Heinze et~al.(2016)Heinze, McWilliams, and
  Meinshausen}]{heinze2016dual}
Heinze, C., McWilliams, B., and Meinshausen, N. (2016), \enquote{Dual-loco:
  Distributing statistical estimation using random projections,} in
  \textit{Artificial Intelligence and Statistics}, pp. 875--883.

\bibitem[{Hosmer~Jr et~al.(2013)Hosmer~Jr, Lemeshow, and
  Sturdivant}]{hosmer2013applied}
Hosmer~Jr, D.~W., Lemeshow, S., and Sturdivant, R.~X. (2013), \textit{Applied
  logistic regression}, vol. 398, John Wiley \& Sons.

\bibitem[{Jordan et~al.(2019)Jordan, Lee, and Yang}]{jordan2018communication}
Jordan, M.~I., Lee, J.~D., and Yang, Y. (2019),
  \enquote{Communication-efficient distributed statistical inference,}
  \textit{Journal of the American Statistical Association}, 114, 668--681.

\bibitem[{Kleiner et~al.(2014)Kleiner, Talwalkar, Sarkar, and
  Jordan}]{kleiner2014scalable}
Kleiner, A., Talwalkar, A., Sarkar, P., and Jordan, M.~I. (2014), \enquote{A
  scalable bootstrap for massive data,} \textit{Journal of the Royal
  Statistical Society: Series B: Statistical Methodology}, 795--816.

\bibitem[{Lee et~al.(2016)Lee, Sun, Sun, Taylor, et~al.}]{lee2016exact}
Lee, J.~D., Sun, D.~L., Sun, Y., Taylor, J.~E., et~al. (2016), \enquote{Exact
  post-selection inference, with application to the lasso,} \textit{Annals of
  Statistics}, 44, 907--927.

\bibitem[{Lee et~al.(2015)Lee, Sun, Liu, and Taylor}]{lee2015communication}
Lee, J.~D., Sun, Y., Liu, Q., and Taylor, J.~E. (2015),
  \enquote{Communication-efficient sparse regression: a one-shot approach,}
  \textit{arXiv preprint arXiv:1503.04337}.

\bibitem[{Lee and Taylor(2014)}]{lee2014exact}
Lee, J.~D. and Taylor, J.~E. (2014), \enquote{Exact post model selection
  inference for marginal screening,} \textit{arXiv preprint arXiv:1402.5596}.

\bibitem[{Lehmann and Casella(2006)}]{lehmann2006theory}
Lehmann, E.~L. and Casella, G. (2006), \textit{Theory of point estimation},
  Springer Science \& Business Media.

\bibitem[{Li et~al.(2012{\natexlab{a}})Li, Peng, Zhang, Zhu,
  et~al.}]{li2012robust}
Li, G., Peng, H., Zhang, J., Zhu, L., et~al. (2012{\natexlab{a}}),
  \enquote{Robust rank correlation based screening,} \textit{The Annals of
  Statistics}, 40, 1846--1877.

\bibitem[{Li et~al.(2012{\natexlab{b}})Li, Zhong, and Zhu}]{li2012feature}
Li, R., Zhong, W., and Zhu, L. (2012{\natexlab{b}}), \enquote{Feature screening
  via distance correlation learning,} \textit{Journal of the American
  Statistical Association}, 107, 1129--1139.

\bibitem[{Li et~al.(2019)Li, Li, Xia, and Xu}]{li2019distributedFeatures}
Li, X., Li, R., Xia, Z., and Xu, C. (2019), \enquote{Distributed feature
  screening via componentwise debiasing,} \textit{arXiv preprint
  arXiv:1903.03810}.

\bibitem[{Li et~al.(2020)Li, Li, Xia, and Xu}]{li2020distributed}
--- (2020), \enquote{Distributed Feature Screening via Componentwise
  Debiasing.} \textit{Journal of Machine Learning Research}, 21, 1--32.

\bibitem[{Liu and Ihler(2014)}]{liu2014distributed}
Liu, Q. and Ihler, A.~T. (2014), \enquote{Distributed estimation, information
  loss and exponential families,} in \textit{Advances in neural information
  processing systems}, pp. 1098--1106.

\bibitem[{Pedregosa et~al.(2011)Pedregosa, Varoquaux, Gramfort, Michel,
  Thirion, Grisel, Blondel, Prettenhofer, Weiss, Dubourg, Vanderplas, Passos,
  Cournapeau, Brucher, Perrot, and Duchesnay}]{scikit-learn}
Pedregosa, F., Varoquaux, G., Gramfort, A., Michel, V., Thirion, B., Grisel,
  O., Blondel, M., Prettenhofer, P., Weiss, R., Dubourg, V., Vanderplas, J.,
  Passos, A., Cournapeau, D., Brucher, M., Perrot, M., and Duchesnay, E.
  (2011), \enquote{Scikit-learn: Machine Learning in {P}ython,} \textit{Journal
  of Machine Learning Research}, 12, 2825--2830.

\bibitem[{Sengupta et~al.(2016)Sengupta, Volgushev, and
  Shao}]{sengupta2016subsampled}
Sengupta, S., Volgushev, S., and Shao, X. (2016), \enquote{A subsampled double
  bootstrap for massive data,} \textit{Journal of the American Statistical
  Association}, 111, 1222--1232.

\bibitem[{Shamir et~al.(2014)Shamir, Srebro, and
  Zhang}]{shamir2014communication}
Shamir, O., Srebro, N., and Zhang, T. (2014), \enquote{Communication-efficient
  distributed optimization using an approximate newton-type method,} in
  \textit{International conference on machine learning}, pp. 1000--1008.

\bibitem[{Shao(1997)}]{shao1997asymptotic}
Shao, J. (1997), \enquote{An asymptotic theory for linear model selection,}
  \textit{Statistica Sinica}, 221--242.

\bibitem[{Smith et~al.(2018)Smith, Forte, Ma, Tak{\'a}{\v{c}}, Jordan, and
  Jaggi}]{smith2018cocoa}
Smith, V., Forte, S., Ma, C., Tak{\'a}{\v{c}}, M., Jordan, M.~I., and Jaggi, M.
  (2018), \enquote{CoCoA: A General Framework for Communication-Efficient
  Distributed Optimization,} \textit{Journal of Machine Learning Research}, 18,
  1--49.

\bibitem[{Taylor and Tibshirani(2015)}]{taylor2015statistical}
Taylor, J. and Tibshirani, R.~J. (2015), \enquote{Statistical learning and
  selective inference,} \textit{Proceedings of the National Academy of
  Sciences}, 112, 7629--7634.

\bibitem[{Tibshirani(1996)}]{tibshirani1996regression}
Tibshirani, R. (1996), \enquote{Regression shrinkage and selection via the
  lasso,} \textit{Journal of the Royal Statistical Society. Series B},
  267--288.

\bibitem[{Volgushev et~al.(2019)Volgushev, Chao, Cheng,
  et~al.}]{volgushev2019distributed}
Volgushev, S., Chao, S.-K., Cheng, G., et~al. (2019), \enquote{Distributed
  inference for quantile regression processes,} \textit{The Annals of
  Statistics}, 47, 1634--1662.

\bibitem[{Wang and Leng(2007)}]{wang2007unified}
Wang, H. and Leng, C. (2007), \enquote{Unified LASSO estimation by least
  squares approximation,} \textit{Journal of the American Statistical
  Association}, 102, 1039--1048.

\bibitem[{Wang et~al.(2007)Wang, Li, and Tsai}]{wang2007tuning}
Wang, H., Li, R., and Tsai, C.-L. (2007), \enquote{Tuning parameter selectors
  for the smoothly clipped absolute deviation method,} \textit{Biometrika}, 94,
  553--568.

\bibitem[{Wang et~al.(2018)Wang, Zhu, and Ma}]{wang2018optimal}
Wang, H., Zhu, R., and Ma, P. (2018), \enquote{Optimal subsampling for large
  sample logistic regression,} \textit{Journal of the American Statistical
  Association}, 113, 829--844.

\bibitem[{Wang et~al.(2017{\natexlab{a}})Wang, Kolar, Srebro, and
  Zhang}]{wang2017efficient}
Wang, J., Kolar, M., Srebro, N., and Zhang, T. (2017{\natexlab{a}}),
  \enquote{Efficient distributed learning with sparsity,} in
  \textit{Proceedings of the 34th International Conference on Machine
  Learning-Volume 70}, JMLR. org, pp. 3636--3645.

\bibitem[{Wang et~al.(2017{\natexlab{b}})Wang, Wang, and
  Srebro}]{pmlr-v65-wang17a}
Wang, J., Wang, W., and Srebro, N. (2017{\natexlab{b}}), \enquote{Memory and
  communication efficient distributed stochastic optimization with minibatch
  prox,} in \textit{Proceedings of the 2017 Conference on Learning Theory},
  eds. Kale, S. and Shamir, O., Amsterdam, Netherlands: PMLR, vol.~65 of
  \textit{Proceedings of Machine Learning Research}, pp. 1882--1919.

\bibitem[{Wang et~al.(2013)Wang, Kim, and Li}]{wang2013calibrating}
Wang, L., Kim, Y., and Li, R. (2013), \enquote{Calibrating non-convex penalized
  regression in ultra-high dimension,} \textit{Annals of statistics}, 41, 2505.

\bibitem[{Yang et~al.(2016)Yang, Mahoney, Saunders, and Sun}]{yang2016feature}
Yang, J., Mahoney, M.~W., Saunders, M., and Sun, Y. (2016),
  \enquote{Feature-distributed sparse regression: a screen-and-clean approach,}
  in \textit{Advances in Neural Information Processing Systems}, pp.
  2712--2720.

\bibitem[{Yu et~al.(2020)Yu, Chao, and Cheng}]{yu2020simultaneous}
Yu, Y., Chao, S.-K., and Cheng, G. (2020), \enquote{Simultaneous Inference for
  Massive Data: Distributed Bootstrap,} \textit{arXiv preprint
  arXiv:2002.08443}.

\bibitem[{Zaharia et~al.(2012)Zaharia, Chowdhury, Das, Dave, Ma, McCauley,
  Franklin, Shenker, and Stoica}]{zaharia2012resilient}
Zaharia, M., Chowdhury, M., Das, T., Dave, A., Ma, J., McCauley, M., Franklin,
  M.~J., Shenker, S., and Stoica, I. (2012), \enquote{Resilient distributed
  datasets: A fault-tolerant abstraction for in-memory cluster computing,} in
  \textit{Proceedings of the 9th USENIX conference on Networked Systems Design
  and Implementation}, USENIX Association, pp. 2--2.

\bibitem[{Zhang(2010)}]{zhang2010nearly}
Zhang, C.-H. (2010), \enquote{Nearly unbiased variable selection under minimax
  concave penalty,} \textit{Annals of Statistics}, 38, 894--942.

\bibitem[{Zhang and Lu(2007)}]{zhang2007adaptive}
Zhang, H.~H. and Lu, W. (2007), \enquote{Adaptive Lasso for Cox's proportional
  hazards model,} \textit{Biometrika}, 94, 691--703.

\bibitem[{Zhang(2004)}]{zhang2004solving}
Zhang, T. (2004), \enquote{Solving large scale linear prediction problems using
  stochastic gradient descent algorithms,} in \textit{Proceedings of the
  twenty-first international conference on Machine learning}, ACM, p. 116.

\bibitem[{Zhang et~al.(2013)Zhang, Duchi, and
  Wainwright}]{zhang2013communication}
Zhang, Y., Duchi, J.~C., and Wainwright, M.~J. (2013),
  \enquote{Communication-efficient algorithms for statistical optimization,}
  \textit{The Journal of Machine Learning Research}, 14, 3321--3363.

\bibitem[{Zhu et~al.(2011)Zhu, Li, Li, and Zhu}]{zhu2011model}
Zhu, L.-P., Li, L., Li, R., and Zhu, L.-X. (2011), \enquote{Model-free feature
  screening for ultrahigh-dimensional data,} \textit{Journal of the American
  Statistical Association}, 106, 1464--1475.

\bibitem[{Zou(2006)}]{zou2006adaptive}
Zou, H. (2006), \enquote{The adaptive lasso and its oracle properties,}
  \textit{Journal of the American Statistical Association}, 101, 1418--1429.

\bibitem[{Zou and Li(2008)}]{zou2008one}
Zou, H. and Li, R. (2008), \enquote{One-step sparse estimates in nonconcave
  penalized likelihood models,} \textit{Annals of statistics}, 36, 1509.

\bibitem[{Zou and Zhang(2009)}]{zou2009adaptive}
Zou, H. and Zhang, H.~H. (2009), \enquote{On the adaptive elastic-net with a
  diverging number of parameters,} \textit{Annals of statistics}, 37, 1733.

\end{thebibliography}

\newpage
\appendix
\section*{Appendix A}
\renewcommand{\theequation}{A.\arabic{equation}}
\setcounter{equation}{0}


\subsection*{Appendix A.1: Proof of Proposition \ref{prop1}}

Note that $\wt \theta - \theta_0$ takes the form
\beq
\wt\theta - \theta_0 = \big\{\sum_k \alpha_k \wh\Sigma_k^{-1}\big\}^{-1}\big\{
\sum_k \alpha_k \wh\Sigma_k^{-1}(\wh\theta_k - \theta_0)\big\}.\nonumber
\eeq
Define $\wh \Sigma_k(\theta) = \{\partial^2\mL_k(\theta)/\partial \theta\partial\theta^\top\}^{-1}$.
In the following section, we denote $\wh\Sigma_k$ by $\wh \Sigma_k(\wh \theta_k)$ to make it  clearer.
By Slutsky's Theorem, to prove (\ref{bias_variance}), it suffices to verify that
\begin{align}
&\sum_k\alpha_k\{\wh\Sigma_k(\wh\theta_k)\}^{-1}\rightarrow_p \Sigma^{-1},\label{Sig_conv}\\
&\sqrt{N}\Big[\sum_k \alpha_k \{\wh\Sigma_k(\wh\theta_k)\}^{-1}(\wh\theta_k - \theta_0)\Big] = V^*(\theta_0) + B^*(\theta_0) ,\label{theta_conv}
\end{align}
where $\cov\{V^*(\theta_0)\} = \Sigma^{-1}$
and $B^*(\theta_0) = O_p(K/\sqrt N)$.
We prove them in the following.

{\sc 1. Proof of (\ref{Sig_conv}).}
Recall that $\wh\theta_k$ is a $\sqrt n$-consistent estimator of $\theta_0$.
This enables us to conduct a Taylor's expansion of $\wh\Sigma_k^{-1}(\wh\theta_k)$ at $\theta_0$, which yields
\begin{align*}
&\wh \Sigma_k^{-1}(\wh \theta_k) - \Sigma_k^{-1} =
\wh \Sigma_k^{-1}(\wh \theta_k) - \wh \Sigma_k^{-1}(\theta_0)
+ \wh \Sigma_k^{-1}(\theta_0)
- \Sigma_k^{-1}\\
&  = \sum_j\frac{\partial^3 \mL_k(\theta)}{\partial\theta_j \partial\theta \partial\theta^\top}\Big|_{\theta = \theta^*}(\theta_j^* - \theta_j)
+ \wh \Sigma_k^{-1}(\theta_0)
- \Sigma_k^{-1}
\end{align*}
where $\theta^*$ lies on the line joining $\theta_0$ and $\wh\theta_k$.
By Condition (C5), we have $ \frac{\partial^3 \mL_k(\theta)}{\partial\theta_j \partial\theta \partial\theta^\top}\Big|_{\theta = \theta^*} = O_p(1)$.
Therefore, the order of the first term is $O_p(1/\sqrt n_k)$.
In addition, we have
$\wh \Sigma_k^{-1}(\theta_0) - \Sigma_k^{-1} = \wh \Sigma_k^{-1}(\theta_0) - E\{\wh \Sigma_k^{-1}(\theta_0)\}
= O_p(n_k^{-1/2})$.
Consequently, it can be derived that $\wh \Sigma_k^{-1}(\wh \theta_k) - \Sigma_k^{-1} = O_p(n_k^{-1/2})$.
Further note that $\alpha_k = n_k/N$ and $\sum_k \alpha_k  =1$.
Then, we have
\beq
\wh\Sigma - \Sigma = \sum_k \alpha_k [\{\wh\Sigma_k(\wh\theta_k)\}^{-1} - \Sigma_k^{-1}] = O_p(n^{-1/2}) = o_p(1).\label{Sig_diff}
\eeq
Hence (\ref{Sig_conv}) is proven.

{\sc 2. Proof of (\ref{theta_conv}).}
Recall that $\wh\theta_k$ is the local minimizer of $\mL_k(\theta)$.
Therefore, it holds that
\begin{align*}
0 = \frac{\partial \mL_k(\theta)}{\partial \theta}\Big|_{\theta = \wh\theta_k} = &\frac{\partial \mL_k(\theta)}{\partial \theta}\Big|_{\theta = \theta_0} + \frac{1}{n_k}\sum_{i\in\mS_k}\frac{\partial^2 \mL(\theta;Z_i)}{\partial \theta\partial \theta^\top}\Big|_{\theta = \theta_0}(\wh\theta_k - \theta_0) \\
&+ \frac{1}{2n_k}\sum_{i\in\mS_k}\sum_{j = 1}^p(\theta^* - \theta_0)^\top
\frac{\partial^3 \mL(\theta;Z_i)}{\partial\theta_j\partial\theta\partial\theta^\top}\Big |_{\theta = \theta^*}(\theta^* - \theta_0),
\end{align*}
where $\theta^*$ lies between $\theta_0$ and $\wh\theta_k$.
By standard arguments,
\begin{align*}
&\frac{\partial \mL_k(\theta)}{\partial\theta}\Big|_{\theta = \theta_0} = O_p(n_k^{-1/2}),\\
& \wh\Sigma_k^{-1}(\theta_0) = E\Big\{
\frac{\partial^2 \mL(\theta; Z_i, i\in \mS_k)}{\partial\theta \partial \theta^\top}\Big\}\Big|_{\theta = \theta_0} + O_p(n_k^{-1/2}) = \Sigma_k^{-1} + O_p(n_k^{-1/2}).\\
& \frac{\partial^3 \mL_k(\theta)}{\partial \theta_j\partial\theta \partial \theta^\top}\Big |_{\theta = \theta^*} = E\Big\{
\frac{\partial^3 \mL(\theta; Z_i, i\in \mS_k)}{\partial\theta \partial \theta^\top\partial \theta_j}\Big|_{\theta = \theta^*}\Big\} + o_p(1).
\end{align*}
Further note that $\wh\theta_k - \theta_0 = O_p(n_k^{-1/2})$.
Then, we have
\beq
\wh\theta_k - \theta_0 = -\Sigma_k\frac{
\partial\mL_k(\theta)}{\partial\theta}\Big |_{\theta = \theta_0}+\frac{B_k(\theta_0)}{n_k} + O_p\Big(\frac{1}{n_k}\Big),\nonumber
\eeq
where $B_k(\theta_0) = O(1)$ is the bias term.
Then, it holds that
\begin{align}
&\sqrt{N}\sum_k \alpha_k \wh\Sigma_k^{-1}(\wh\theta_k)(\wh\theta_k - \theta_0) \nonumber\\
&=
\sqrt{N}\sum_k \alpha_k \Sigma_k^{-1}\Big\{-\Sigma_k\frac{\partial\mL_k(\theta)}{\partial\theta}\Big|_{\theta = \theta_0}
+\frac{B_k(\theta_0)}{n_k} + O_p(n_k^{-1})\Big\} \nonumber\\
&+
\sqrt{N}\sum_k \alpha_k \{\wh\Sigma_k^{-1}(\wh\theta_k) - \Sigma_k^{-1}\}(\wh\theta_k - \theta_0)\nonumber\\
& = -\frac{1}{\sqrt N}\sum_k n_k\frac{\partial\mL_k(\theta)}{\partial\theta}\Big|_{\theta = \theta_0}
+
\frac{1}{\sqrt{N}}\sum_k\Sigma_k^{-1}B_k(\theta_0)
+ O_p\Big(\frac{K}{\sqrt N}\Big),\label{theta_expan}
\end{align}
where the second equation is implied by
$\sqrt{N}\sum_k \alpha_k \{\wh\Sigma_k^{-1}(\wh\theta_k) - \Sigma_k^{-1}\}(\wh\theta_k - \theta_0)  = O_p(K/\sqrt N)$.
By condition (C4), it can be concluded that
$\cov\{\frac{1}{\sqrt N} \sum_kn_k\frac{\partial\mL_k(\theta)}{\partial\theta}|_{\theta = \theta_0}\} = \Sigma^{-1}$.
Consequently, (\ref{theta_conv}) can be proven.

\subsection*{Appendix A.2: Proof of Theorem \ref{thm1}}

By Slutsky's Theorem, the asymptotic normality is directly implied by
(\ref{Sig_conv}) and (\ref{theta_conv}) with
$V^*(\theta_0)\rightarrow_d N(0,\Sigma^{-1})$
and $B^*(\theta_0) = o_p(1)$.
First, by Condition (C6) and the Lyapunov central limit theorem,
we have $V^*(\theta_0)\rightarrow_d N(0,\Sigma^{-1})$.
Next, by
the condition that $n\gg \sqrt N$, we have $K\gg \sqrt N$,
and thus, $B^*(\theta_0) = o_p(1)$.

\subsection*{Appendix A.3: Proof of Theorem \ref{thm_twlse}}

Note that we have
\beq
\wt\theta^{(2)} = \Big(\sum_k \alpha_k \wh\Sigma_k^{(2)-1}\Big)^{-1}\Big(
\sum_k\wh\Sigma_k^{(2)-1}\wh\theta_k^{(2)}\Big).\nonumber
\eeq
It suffices to show that,
\begin{align}
&\sum_k\alpha_k\{\wh\Sigma_k^{(2)}\}^{-1}\rightarrow_p \Sigma^{-1},\label{Sig_conv2}\\
&\sqrt{N}\Big[\sum_k \alpha_k \{\wh\Sigma_k^{(2)}\}^{-1}(\wh\theta_k^{(2)} - \theta_0)\Big] = V_2^*(\theta_0) + B_2^*(\theta_0) ,\label{theta_conv2}\\
& \sqrt{N}(\wt\theta^{(2)} - \theta_0)\rightarrow_d N(\zero, \Sigma),\label{normality2}
\end{align}
where $\cov\{V_2^*(\theta_0)\} = \Sigma^{-1}$
and $B_2^*(\theta_0) = O_p(1/(n^2\sqrt N))$.
We prove them in the following three steps.

{\sc Step 1. $\wh \theta_k^{(2)} - \theta_0 = O_p(n^{-1/2})$.}

Note that we have
\begin{align*}
\wh \theta_k^{(2)} = \wt\theta - \Big(\frac{\partial^2 \mL_k(\wt\theta )}{\partial\theta\partial\theta^\top}\Big)^{-1}\frac{\partial \mL_k(\wt\theta)}{\partial\theta}.
\end{align*}
By performing a Taylor's expansion of $\partial\mL_k(\wt\theta)/\partial\theta$
at $\theta_0$, we could obtain,
\begin{align*}
\frac{\partial\mL_k(\wt\theta)}{\partial\theta} &=
\frac{\partial\mL_k(\theta_0)}{\partial \theta} +
\frac{\partial^2\mL_k(\ol\theta)}{\partial \theta\partial\theta^\top}(\wt\theta
-\theta_0)\\
& = \frac{\partial\mL_k(\theta_0)}{\partial \theta}
+ \Big\{\frac{\partial^2\mL_k(\ol\theta)}{\partial\theta\partial\theta^\top} -
\frac{\partial \mL_k(\wt\theta)}{\partial\theta\partial\theta^\top}\Big\}(\wt\theta - \theta_0) + \frac{\partial^2\mL_k(\wt\theta)}{\partial\theta\partial\theta}({\wt\theta - \theta_0}),
\end{align*}
where $\ol\theta$ lies on the line joining $\wt\theta$ and $\theta_0$.
As a result, we have
\begin{align}
\wh \theta_k^{(2)} &= \theta_0-\Big(
\frac{\partial^2\mL_k(\wt\theta)}{\partial\theta\partial
\theta^\top}\Big)^{-1}\frac{\partial \mL_k(\theta_0)}{\partial\theta}
- \Big(
\frac{\partial^2\mL_k(\wt\theta)}{\partial\theta\partial
\theta^\top}\Big)^{-1}
\Big\{\frac{\partial^2\mL_k(\ol\theta)}{\partial\theta\partial\theta^\top} -
\frac{\partial \mL_k(\wt\theta)}{\partial\theta\partial\theta^\top}\Big\}(\wt\theta - \theta_0)
\nonumber\\
&\defeq \theta_0 - \Delta_{k1} - \Delta_{k2} \label{theta2}
\end{align}
Using Proposition \ref{prop1} we have $\Delta_{k1} = O_p(n^{-1/2})$.
Then it suffices to derive the rate of $\Delta_{k2}$.

By using Taylor's expansion again we have
\begin{align*}
\frac{\partial^2\mL_k(\ol\theta)}{\partial\theta\partial\theta^\top} -
\frac{\partial \mL_k(\wt\theta)}{\partial\theta\partial\theta^\top} =
\sum_j\frac{\partial^3 \mL_k(\ol \theta^{(2)})}{\partial\theta\partial\theta^\top\partial\theta_j}(\ol\theta_j - \wt\theta_{j}),
\end{align*}
where $\ol\theta^{(2)}$ lies on the line joining $\ol\theta$ and
and $\wt\theta$.
By Proposition \ref{prop1} we can obtain
${\partial^3 \mL_k(\ol \theta^{(2)})}/{\partial\theta\partial\theta^\top\partial\theta_j}\rightarrow_p
E\{{\partial^3 \mL_k(\theta_0)}/{\partial\theta\partial\theta^\top\partial\theta_j}\}
\defeq \mL_{kj}^{(3)}$
and $\partial^2\mL_k(\wt\theta)/{\partial\theta\partial
\theta^\top}\rightarrow_p \Sigma_k^{-1}$.
By Proposition \ref{prop1}, we have
\begin{align}
&\Big(
\frac{\partial^2\mL_k(\wt\theta)}{\partial\theta\partial
\theta^\top}\Big)^{-1}
\Big\{\frac{\partial^2\mL_k(\ol\theta)}{\partial\theta\partial\theta^\top} -
\frac{\partial \mL_k(\wt\theta)}{\partial\theta\partial\theta^\top}\Big\}(\wt\theta - \theta_0)\nonumber\\
& = \Sigma_k\sum_j\mL_{kj}^{(3)}\Big\{V^{*(j)}(\theta_0) + B^{*(j)}(\theta_0)\Big\}\{1+o_p(1)\},\label{Delta_2}
\end{align}
where $V^{*(j)}(\theta_0) = \{V(\theta_0)B_j(\theta_0) + B(\theta_0)V_j(\theta_0)\}/(n\sqrt{N}) =
O_p(1/(n\sqrt N))$,
and $B^{*(j)}(\theta_0) = B(\theta_0)B_j(\theta_0) = O_p(n^{-2})$.

{\sc Step 2. Proof of (\ref{Sig_conv2}).}

Following the same procedure of proving (\ref{Sig_conv}) and using $\wh\theta_k^{(2)} - \theta_0 = O_p(n^{-1/2})$ proved
in the first step we could obtain the result.

{\sc Step 3. Proof of (\ref{theta_conv2}).}

By (\ref{theta2}) and (\ref{Delta_2}) we have
\begin{align*}
&\sum_k \alpha_k \{\wh\Sigma_k^{(2)}\}^{-1}(\wh\theta_k^{(2)} - \theta_0) = \sum_k\alpha_k\Big[- \frac{\partial \mL_k(\theta_0)}{\partial\theta}
- \Big\{\frac{\partial^2\mL_k(\ol\theta)}{\partial\theta\partial\theta^\top} -
\frac{\partial \mL_k(\wt\theta)}{\partial\theta\partial\theta^\top}\Big\}(\wt\theta - \theta_0)\Big]\\
& = -\sum_{k}\alpha_k \frac{\partial \mL_k(\theta_0)}{\partial\theta}-
\sum_k \alpha_k\sum_j\mL_{kj}^{(3)}\Big\{V^{*(j)}(\theta_0) + B^{*(j)}(\theta_0)\Big\}\{1+o_p(1)\}
\end{align*}
Define
\begin{align*}
&V_{2}^*(\theta_0) \defeq  -\sqrt N \sum_{k}\alpha_k \frac{\partial \mL_k(\theta_0)}{\partial\theta}\\
&V_{22}(\theta_0) \defeq -\sqrt N \sum_k \alpha_k\sum_j\mL_{kj}^{(3)}V^{*(j)}(\theta_0)\{1+o_p(1)\} \\
&B_2^*(\theta_0) \defeq - \sqrt N \sum_k \alpha_k\sum_j\mL_{kj}^{(3)}B^{*(j)}(\theta_0)\{1+o_p(1)\}.
\end{align*}
Then we have $\cov\{V_{2}^*(\theta_0)\} = \Sigma^{-1}$,
$V_{22}(\theta_0) = o_p(1)$,
and $B_2^*(\theta_0) = O_p(n^{-2}N^{1/2})$
by {\sc Step 2}.

{\sc Step 4. Proof of (\ref{normality2}).}

by Condition (C6) and the Lyapunov central limit theorem, we have $V_2^*(\theta_0)\rightarrow_d N(0, \Sigma^{-1})$.
Under the condition that $n^2/N^{1/2}\rightarrow\infty$, we conclude that
$B_2^*(\theta_0) = o_p(1)$.
Using the Slutsky's Theorem we could obtain the result.

\subsection*{Appendix A.4: Proof of Proposition \ref{mwlse}}

The result can be proved by induction method.
By Theorem \ref{thm_twlse}, the bias term of $\wt\theta^{(2)}$ reduces from $O_p(n^{-1})$ to $O_p(n^{-2})$ when we take one more round of iteration.
Applying the proof of Theorem \ref{thm_twlse} sequentially for $m$ times,
we can show that the bias term of $\wt\theta^{(m)}$ is in the order of $O_p(n^{-m})$.
If $n^{-m}\ll N^{-1/2}$, i.e., $m\gg \log N/\log n$, then the bias term can be ignored.
As a result, following the same proof procedure of Theorem \ref{thm_twlse},
we can obtain the result.

\section*{Appendix B}
\renewcommand{\theequation}{B.\arabic{equation}}
\setcounter{equation}{0}

\subsection*{Appendix B.1: Proof of Theorem \ref{consistency}}

{\sc 1. Proof of $\sqrt N$-consistency.}

Note that the objective function $Q_\lambda(\theta)$ in (\ref{aLasso})
is a strictly convex function.
Then, the local minimizer is also a global minimizer.
To establish $\sqrt N$-consistency results,
it suffices to verify the following result \citep{fan2001variable},
i.e.,
for an arbitrarily small $\epsilon > 0$,
there exists a sufficiently large constant $C$ such that
\beq
\lim_N\inf P\Big\{\inf_{u\in\mR^p:\|u\| = C}Q_{\lambda}(\theta_0 + N^{-1/2}u)>Q(\theta_0)\Big\}>1-\epsilon.\label{aLasso_consistent}
\eeq
Let $u = (u_1,\cdots, u_p)^\top$
and $\wh\Delta_N = \sum_k\alpha_k\wh\Sigma_k^{-1}(\wh\theta_k)\big\{\sqrt{N}(\theta_0 - \wh\theta_k)\big\}$,
Then, we have
\begin{align}
N&\Big\{Q_\lambda(\theta_0 + N^{-1/2}u) - Q_\lambda(\theta_0)\Big\}\nonumber\\
&=u^\top \wh \Sigma^{-1} u + 2u^\top\wh\Delta_N+ N\sum_{j = 1}^p \lambda_j|\theta_{0j}+ N^{-1/2}u_j|-N\sum_{j = 1}^p \lambda_j |\theta_{0j}|\nonumber\\
&\ge u^\top \wh \Sigma^{-1} u + 2u^\top\wh\Delta_N + N \sum_{j = 1}^{d_0}\lambda_j \big(|\theta_{0j} + N^{-1/2}u_j| - |\theta_{0j}|\big)\nonumber\\
&\ge  u^\top \wh \Sigma^{-1} u + 2u^\top\wh\Delta_N -d_0(\sqrt N a_N)\|u\|.\label{Q_diff}
\end{align}
where the second equality holds because we assume $\theta_{0j} = 0$ for $j > d_0$.
Further note that we assume that $\sqrt N a_N\rightarrow_p 0.$
Consequently, the last term (\ref{Q_diff}) is $o_p(1)$.
Next, note that the first term of
(\ref{Q_diff}) is lower bounded by $\lambda_{\max}^{-1}(\wh\Sigma)C^2$ because $\|u\| = C$.
By (\ref{Sig_diff}), we have
$\lambda_{\max}(\wh\Sigma)\rightarrow_p \lambda_{\max}(\Sigma)$.
Consequently, with probability tending to 1, we have the first term uniformly larger
than $0.5\lambda_{\max}^{-1}(\Sigma)C^2$,
which is positive due to Condition (C4).
In addition,
by $K/\sqrt N\rightarrow 0$, we have
$\wh \Delta_N = O_p(1)$.
Consequently, as long as $C$ is sufficiently large, the first term will dominate the last two terms.
Then, the result of (\ref{aLasso_consistent}) is proven.

{\sc 2. Proof of Selection Consistency.}

It suffices to verify that $P(\wt\theta_{\lambda,j} = 0)\rightarrow 1$ for any $d_0<j\le p$.
Note that $Q_\lambda(\theta)$ can be rewritten as
\beq
Q_\lambda(\theta) = (\theta - \wt\theta)^\top \wh \Sigma^{-1}(\theta - \wt\theta)+ \sum_j\lambda_j|\theta_j|+C,\nonumber
\eeq
where $C$ is a constant.
Define $\wh\Omega = \wh\Sigma^{-1}$, and $\wh\Omega^{(j)}$ denotes the $j$th row of the matrix $\wh\Omega$.
If $\wt\theta_{\lambda,j}\ne 0$ for some $j>d_0$, then the partial derivative can be calculated as
\beq
\sqrt N \frac{\partial Q_\lambda(\theta)}{\partial \theta_j}\Big|_{\theta = \wt\theta_\lambda} =
2 \wh\Omega^{(j)\top} \sqrt N(\wt \theta_\lambda - \wt\theta) + \sqrt N\lambda_j \sign(\wt\theta_{\lambda,j}).\label{Q_partial}
\eeq
Note that $\wh\Omega\rightarrow_p \Sigma^{-1}$ and $\sqrt N(\wt\theta_\lambda - \wt\theta)
= \sqrt N(\wt\theta_\lambda - \theta_0) - \sqrt N(\wt\theta - \theta_0) = O_p(1)$,
by (\ref{Sig_conv}), Theorem \ref{thm1},
and Theorem \ref{consistency} (a).
Consequently, the first term (\ref{Q_partial}) is $O_p(1)$.
Next, by this condition, we know that
$\sqrt N\lambda_j \ge \sqrt Nb_\lambda \rightarrow\infty$ for $j >d_0$.
Because $\wt\theta_{\lambda,j}\ne 0$,  we have $\sign(\wt\theta_{\lambda,j}) = 1$ or -1; thus, the second term (\ref{Q_partial}) goes to infinity.
Obviously, the equation will not be equal to zero.
This implies $P(\wt\theta_{\lambda,j} = 0)\rightarrow 1$ as a result.

\subsection*{Appendix B.2: Proof of Theorem \ref{oracle}}

We first rewrite the asymptotic covariance $\Sigma$ into the following block matrix form:
\begin{align*}
\Sigma = \left(
\begin{array}{cc}
  \Sigma_{11} & \Sigma_{12}  \\
  \Sigma_{21} & \Sigma_{22}
\end{array}\right),
\end{align*}
where $\Sigma_{11}\in\mR^{d_0\times d_0}$.
Similarly, we partition its inverse matrix $\Omega$ into 4 corresponding parts,
$\Omega = (\Omega_{11},\Omega_{12};\Omega_{21},\Omega_{22})$.
By Theorem \ref{consistency},
with probability tending to 1,
we have $\wt\theta_{\lambda}^{(-\mM_T)} = 0$.
Therefore, $\wt\theta_{\lambda}^{(\mM_T)}$ should be the global minimizer of the objective function,
\begin{align}
Q_{\lambda,0}(\theta^{(\mM_T)}) = &
(\theta^{(\mM_T)} - \wt\theta^{(\mM_T)})^\top\wh\Omega_{11}(\theta^{(\mM_T)} - \wt\theta^{(\mM_T)})
-2(\theta^{(\mM_T)} - \wt\theta^{(\mM_T)})^\top\wh\Omega_{12}\wt\theta^{(-\mM_T)}\nonumber\\
&+
\wt\theta^{(-\mM_T)\top}\wh\Omega_{22}\wt\theta^{(-\mM_T)} + \sum_{j = 1}^{d_0}\lambda_j|\theta_j|\nonumber
\end{align}
By Theorem \ref{consistency}, it can be concluded that with probability tending to 1,
$\wt\theta_\lambda^{(\mM_T)}$ should be nonzero (otherwise, the $\sqrt N$-consistency result in Theorem \ref{consistency} will not hold).
As a result,
The partial derivative $\partial Q_\lambda(\theta)/\partial \theta_j$ should exist for $1\le j\le d_0$,
which yields
\begin{align}
0 = \frac{1}{2}\frac{\partial Q_{\lambda,0}(\theta^{(\mM_T)})}{\partial \theta^{(\mM_T)}}\Big|_{\theta^{(\mM_T)} = \wt\theta_{\lambda}^{(\mM_T)}} =
\wh\Omega_{11}(\wt\theta_\lambda^{(\mM_T)} - \wt\theta^{(\mM_T)}) - \wh\Omega_{12}
\wt\theta^{(-\mM_T)} + D(\wt\theta_{\lambda}^{(\mM_T)}).\label{Q_deriv}
\end{align}
where $D(\wt\theta_{\lambda}^{(\mM_T)})$ is a $d_0$-dimensional vector, with its $j$th component given by $0.5\lambda_j\sign(\wt\theta_{\lambda,j})$.
By (\ref{Q_deriv}), it can be derived that
\begin{align}
\sqrt N(\wt\theta_{\lambda}^{(\mM_T)} - \theta_0^{(\mM_T)}) &=
\sqrt N(\wt\theta^{(\mM_T)} - \theta_0^{(\mM_T)}) + \wh\Omega_{11}^{-1}\wh\Omega_{12}(\sqrt N \wt\theta^{(-\mM_T)})-\wh\Omega_{11}^{-1}\sqrt N D(\wt\theta_{\lambda}^{(\mM_T)})\nonumber\\
& = \sqrt N(\wt\theta^{(\mM_T)} - \theta_0^{(\mM_T)}) + \Omega_{11}^{-1}\Omega_{12}(\sqrt N\wt\theta^{(-\mM_T)}) + o_p(1),
\end{align}
where the second equality follows because
$\sqrt N\wt\theta^{(-\mM_T)} = O_p(1)$ by Theorem \ref{thm1},
$\wh\Omega_{11}\rightarrow_p\Omega_{11}$ and
$\wh\Omega_{12}\rightarrow_p\Omega_{12}$ by (\ref{Sig_conv}),
and $\sqrt N\lambda_j = o_p(1)$ ($1\le j\le d_0$) by Theorem \ref{consistency}.
Furthermore, by the matrix inverse formula,
it holds that $\Omega_{11}^{-1}\Omega_{12} = -\Sigma_{12}\Sigma_{22}^{-1}$.
Consequently, we have
\beq
\sqrt N(\wt \theta_{\lambda}^{(\mM_T)} - \theta_0^{(\mM_T)})
= \sqrt N (\wt\theta^{(\mM_T)} - \theta_0^{(\mM_T)}) - \Sigma_{12}\Sigma_{22}^{-1}(\sqrt N \wt\theta^{(-\mM_T)}) + o_p(1).\nonumber
\eeq
By Theorem \ref{thm1}, we have that the above is asymptotically normal with a mean of 0,
and the inverse asymptotic covariance matrix is given by
$(\Sigma_{11} - \Sigma_{12}\Sigma_{22}^{-1}\Sigma_{21})^{-1} = \Omega_{11}$.
By condition (C6), we have $\Omega_{11} = \Omega_{\mM_T}$.
Consequently, the estimator $\wt\theta_{\lambda}^{(\mM)}$
shares the same asymptotic distribution with the oracle estimator $\wh\theta_{\mM_T}^{(\mM_T)}$.

\subsection*{Appendix B.3: Proof of Theorem \ref{thm_dbic}}

To establish the selection consistency property of the DBIC,
we consider the following two cases for any $\mM_\lambda \ne \mM_T$.
The first case is the underfitted case, and the second case is the overfitted case.

{\sc 1. Underfitted Model.}
Note that $\lambda_N$ satisfies the condition in Theorem \ref{consistency}.
Consequently, we have that $\wt\theta_{\lambda_N}$ is $\sqrt N$-consistent.
We thus also have $\dbic_{\lambda_N} = o_p(1)$.
For $\mM\not\supset \mM_T$, it can be derived that
\begin{align*}
\dbic_\lambda &= (\wt\theta_{\lambda} - \wt\theta)^\top \wh\Sigma^{-1}(\wt\theta_\lambda - \wt\theta) + df_\lambda (\log N)/N\\
&\ge (\wt\theta_\lambda - \wt\theta)^\top \wh\Sigma^{-1}(\wt\theta_\lambda - \wt\theta)^\top
\end{align*}
Define $\wt\theta_{\mM} = \arg\min_{\theta\in\mR^p:\theta_j = 0,\forall j\not\in\mM}(\theta - \wt\theta)^\top\wh\Sigma^{-1}(\theta - \wt\theta)$ as the unpenalized estimator given the model identified by $\mM$.
Consequently, by definition, we should have
\begin{align*}
&(\wt\theta_\lambda - \wt\theta)^\top \wh\Sigma^{-1}(\wt\theta_\lambda - \wt\theta)^\top \ge
(\wt\theta_{\mM_\lambda} - \wt\theta)^\top\wh\Sigma^{-1}(\wt\theta_{\mM_{\lambda}} - \wt\theta)\\
& \ge \min_{\mM\not\supset\mM_T} (\wt\theta_{\mM} - \wt\theta)^\top
\wh\Sigma^{-1}(\wt\theta_{\mM} - \wt\theta)\rightarrow_p
 \min_{\mM\not\supset\mM_T} (\wt\theta_{\mM} - \wt\theta)^\top
\Sigma^{-1}(\wt\theta_{\mM} - \wt\theta),
\end{align*}
where the last convergence is due to (\ref{Sig_conv}).
Because $\Sigma$ is positive definite by Condition (C4),
we have $ \min_{\mM\not\supset\mM_T} (\wt\theta_{\mM} - \wt\theta)^\top
\Sigma^{-1}(\wt\theta_{\mM} - \wt\theta)>0$ with probability tending to 1.
One could then conclude immediately that
$P(\inf_{\lambda\in\mR_{-}}\dbic_\lambda > \dbic_{\lambda_N})\rightarrow 1$.

{\sc 2. Overfitted Model.}
We next consider the overfitted model.
In contrast, let $\lambda$ be an arbitrary tuning parameter that over selects the parameters.
We then have $df_\lambda - d_0 \ge 1$.
It can then be concluded that $N(\dbic_\lambda - \dbic_{\lambda_N}) = $
\begin{align}
&N(\wt\theta_\lambda - \wt\theta)^\top \wh\Sigma^{-1} (\wt\theta_\lambda - \wt\theta)
-N(\wt\theta_{\lambda_N} - \wt\theta)^\top \wh\Sigma^{-1} (\wt\theta_{\lambda_N} - \wt\theta)
+ (df_\lambda - d_0)\log N\nonumber\\
&\ge N(\wt\theta_{\mM_\lambda} - \wt\theta)^\top\wh\Sigma^{-1}
(\wt\theta_{\mM_\lambda} - \wt\theta)-
N(\wt\theta_{\lambda_N} - \wt\theta)^\top \wh\Sigma^{-1}
(\wt\theta_{\lambda_N} - \wt\theta)+\log N\nonumber \\
&\ge \inf_{\mM\supset \mM_T} N(\wt\theta_{\mM} - \wt\theta)^\top \wh\Sigma^{-1}
(\wt\theta_{\mM} - \wt\theta) - N(\wt\theta_{\lambda_N} - \wt\theta)^\top \wh\Sigma^{-1}
(\wt\theta_{\lambda_N} - \wt\theta) + \log N.\label{theta_diff}
\end{align}
First note that for $\mM\supset \mM_T$, we
have that $\wt\theta_{\mM}$ is $\sqrt N$-consistent.
As a result, the first term of (\ref{theta_diff})
is $O_p(1)$.
Similarly, by Theorem \ref{consistency},
$\wt\theta_{\lambda_N}$ is $\sqrt N$-consistent,
Thus, the second term (\ref{theta_diff}) is also $O_p(1)$.
As a result, (\ref{theta_diff}) diverges to infinity as $N\rightarrow \infty$.
This implies that $P(\inf_{\lambda \in\mR_+}\dbic_{\lambda} > \dbic_{\lambda_N})\rightarrow 1$.
This completes the proof.

\subsection*{Appendix B.4: Comparison of the Shrinkage Estimations}

One could consider an alternative approach to conduct variable selection.
Based on the asymptotic theory given in Theorem \ref{thm1},
we could perform simultaneous hypothesis testing for the variable selections.
Specifically,
for each $1\le j\le p$, we could construct a 95\% confidence interval for $\theta_{0j}$
as
CI$_j = (\wt\theta_j - z_{0.975}\wt{\mbox{SE}}_j,
\wt\theta_j + z_{0.975}\wt{\mbox{SE}}_j)$,
where $\wt{\mbox{SE}}_j$ is the $j$th diagonal element of $\wh \Sigma = (\sum_k \alpha_k \wh \Sigma_k^{-1})^{-1}$,
and $z_\alpha$ is the $\alpha$th quantile of a standard normal distribution.
As a result, if $0\not\in \mbox{CI}_j$,
we could identify the $j$th variable as an important variable.

We compare the finite sample performance of the above variable selection method with the proposed shrinkage estimation method for the logistic regression model.
We set $p = 30$, and for $1\le j\le 15$, we
specify $\theta_{0j} = U_1\sign(U_2)$,
where $U_1$ and $U_2$ are generated from
uniform distribution $U[0.5, 2]$
and $U[-0.2, 0.8]$ respectively.
In addition, for $15<j\le p$, we set $\theta_{0j} = 0$.
We also denote the percentage for covering true model
and identifying the true model as CR and CM respectively.
As shown in Table \ref{tab_shrinkage}, both methods
are able to cover the true model with high accuracy.
However, the hypothesis testing method tends to over select the important features while the SDLSA method could still guarantee the variable selection consistency property.

\subsection*{Appendix B.5: Interleave with Bootstrap Methods}

An alternative way to deal with statistical inference for massive dataset
is to use bootstrap methods.
Recently, a surge of researches have investigated related methodologies.
For instance, \cite{kleiner2014scalable} and
\cite{sengupta2016subsampled} propose to use subsample and resample methods
to conduct statistical inference to reduce computational cost of traditional
bootstrap methods.
\cite{volgushev2019distributed}
and \cite{yu2020simultaneous} further discuss how to incorporate the
bootstrap methods into distributed learning algorithms.
Particularly, we note that
if we have a large number of workers, we could adopt the idea of
\cite{volgushev2019distributed} to transmit estimators $\wh\theta_k$
to the master and estimate the asymptotic covariance $\wh \Sigma$
as
$\wh \Sigma = K^{-1}\sum_{k = 1}^K (\wh \theta_k - \ol\theta)(\wh \theta_k - \ol\theta)^\top$,
where $\ol \theta = K^{-1}\sum_{k = 1}^K \wh \theta_k$.
The advantage here is to avoid the communication of the second
order derivative of $\mL_k(\theta)$.
However, the estimation requires a large number of workers to be valid
and in addition, the estimated asymptotic covariance is for the one-shot
estimator.
Hence, the methodology cannot be directly applied especially
for our heterogenous data setting.

As an alternative, we can interleave the bootstrap methods
on workers for the estimation of the covariance $\wh \Sigma_k$.
The method can be useful when it is hard to obtain an analytical form
of the asymptotic covariance $\wh \Sigma_k$.
To show that it is a possible direction  for future study,
we conduct a simulation study for the logistic regression model.
Specifically, we use the subsampled double bootstrap (SDB) estimation method
on workers to obtain the covariance estimator $\wh \Sigma_{k}^{boot}$.
Next, we substitute $\wh \Sigma_k^{boot}$
to $\wh \Sigma_k$ in (\ref{theta_tilde})
to obtain the bootstraped-WLSE (BWLSE) estimator.
We set the bootstrap sample size $b = n^{\gamma}$ with $\gamma = 0.9$
and the resampling times $R = n_k$.
The estimation REE with respect to global estimator is presented in Table \ref{bootstrap}.
Under the i.i.d data setting, all the three methods are comparable when $N = 20,000$
and $K = 10$.
However, under the heterogenous data setting,
one could observe that the bootstrap method is better than the OS method
and is comparable to the DLSA method.

\subsection*{Appendix B.6: Supplementary results for simulation studies}

\begin{table}%
  \caption{Simulation results for Example 1 (Setting 1) with 500 replications.  The numerical
    performances are evaluated for different sample sizes $N$ ($\times 10^3$) and numbers
    of workers $K$.
    The REE is reported for all estimators.
    For the CSL, DLSA, TDLSA method, the CP is further reported in parentheses.
    Finally, the
    MS and CM are reported for the SDLSA method to evaluate the variable selection
    accuracy.  }\label{tab_linear_iid}
\begin{center}
\begin{tabular}{lcccccccccc}
\toprule
 Est.  & $\theta_1$ & $\theta_2$ & $\theta_3$ & $\theta_4$ & $\theta_5$ & $\theta_6$ & $\theta_7$ & $\theta_8$ & MS   & CM   \\
  \midrule
  \multicolumn{11}{c}{$N= 20$, $K=10$}                                                                                                   \\

 OS    & 1.00       & 1.00       & 1.00       & 1.00       & 1.00       & 1.00       & 1.00       & 0.99       &      &      \\
 CSL   & 0.96       & 0.93       & 0.95       & 0.93       & 0.95       & 0.96       & 0.93       & 0.96       &      &      \\
       & (94.0)     & (94.0)     & (94.8)     & (92.8)     & (92.8)     & (93.2)     & (90.8)     & (90.4)     &      &      \\
 DLSA  & 1.00       & 1.00       & 1.00       & 1.00       & 1.00       & 1.00       & 1.00       & 1.00       &      &      \\
       & (96.4)     & (96.0)     & (96.4)     & (94.6)     & (94.6)     & (94.4)     & (94.0)     & (93.8)     &      &      \\
 TDLSA & 1.00       & 1.00       & 1.00       & 1.00       & 1.00       & 1.00       & 1.00       & 1.00       &      &      \\
       & (96.4)     & (96.0)     & (96.4)     & (94.6)     & (94.6)     & (94.4)     & (94.0)     & (93.8)     &      &      \\
 SDLSA & 1.03       & -          & -          & 1.07       & -          & -          & 1.08       & -          & 3.01 & 1.00 \\

  \midrule
  \multicolumn{11}{c}{$N= 100$, $K=20$}                                                                                                   \\
  OS    & 1.00       & 1.00       & 1.00       & 1.00       & 1.00       & 1.00       & 1.00       & 1.00       &      &      \\
 CSL   & 0.95       & 0.92       & 0.93       & 0.94       & 0.95       & 0.93       & 0.92       & 0.96       &      &      \\
       & (93.8)     & (92.2)     & (91.4)     & (92.6)     & (92.2)     & (92.2)     & (92.0)     & (94.6)     &      &      \\
 DLSA  & 1.00       & 1.00       & 1.00       & 1.00       & 1.00       & 1.00       & 1.00       & 1.00       &      &      \\
       & (95.4)     & (94.8)     & (94.8)     & (95.6)     & (95.2)     & (94.0)     & (94.6)     & (95.2)     &      &      \\
 TDLSA & 1.00       & 1.00       & 1.00       & 1.00       & 1.00       & 1.00       & 1.00       & 1.00       &      &      \\
       & (95.4)     & (94.8)     & (94.8)     & (95.6)     & (95.2)     & (94.0)     & (94.6)     & (95.2)     &      &      \\
 SDLSA & 1.03       & -          & -          & 1.05       & -          & -          & 1.08       & -          & 3.01 & 1.00 \\
\bottomrule
\end{tabular}
\end{center}
\end{table}

\begin{table}
  \caption{Simulation results for Example 2 (Setting 1) with 500 replications.  The numerical
    performances are evaluated for different sample sizes $N$ ($\times 10^3$) and numbers
    of workers $K$.
    The REE is reported for all estimators.
    For the CSL, DLSA, TDLSA method, the CP is further reported in parentheses.
    Finally, the
    MS and CM are reported for the SDLSA method to evaluate the variable selection
    accuracy.  }\label{tab_logistic_iid}
\begin{center}
\begin{tabular}{lcccccccccc}
\toprule
 Est.  & $\theta_1$ & $\theta_2$ & $\theta_3$ & $\theta_4$ & $\theta_5$ & $\theta_6$ & $\theta_7$ & $\theta_8$ & MS   & CM   \\
  \midrule
  \multicolumn{11}{c}{$N= 20$, $K=10$}                                                                                                   \\

 OS    & 0.84       & 0.98       & 0.99       & 0.90       & 0.98       & 0.99       & 0.88       & 0.99       &      &      \\
 CSL   & 0.88       & 0.98       & 0.94       & 0.91       & 0.95       & 0.94       & 0.86       & 0.96       &      &      \\
       & (91.0)     & (93.6)     & (93.0)     & (93.8)     & (92.0)     & (92.4)     & (90.2)     & (92.0)     &      &      \\
 DLSA  & 0.93       & 1.01       & 1.01       & 0.96       & 1.01       & 1.01       & 0.94       & 1.01       &      &      \\
       & (92.2)     & (95.6)     & (94.8)     & (94.6)     & (94.2)     & (95.0)     & (92.8)     & (93.2)     &      &      \\
 TDLSA & 1.00       & 1.00       & 1.00       & 1.00       & 1.00       & 1.00       & 1.00       & 1.00       &      &      \\
       & (94.8)     & (95.2)     & (95.0)     & (95.4)     & (93.8)     & (95.0)     & (95.2)     & (93.4)     &      &      \\
 SDLSA & 1.00       & -          & -          & 1.01       & -          & -          & 1.03       & -          & 3.01 & 1.00 \\
  \midrule
  \multicolumn{11}{c}{$N= 100$, $K=20$}                                                                                                   \\

  OS    & 0.87       & 1.00       & 1.00       & 0.91       & 1.00       & 1.00       & 0.92       & 0.99       &      &      \\
 CSL   & 0.88       & 0.98       & 0.91       & 0.93       & 0.94       & 0.93       & 0.90       & 0.96       &      &      \\
       & (90.6)     & (94.0)     & (91.8)     & (92.4)     & (94.8)     & (93.2)     & (91.8)     & (94.8)     &      &      \\
 DLSA  & 0.92       & 1.00       & 1.00       & 0.98       & 1.00       & 1.00       & 0.92       & 1.00       &      &      \\
       & (92.0)     & (95.4)     & (96.8)     & (93.8)     & (97.2)     & (95.8)     & (92.2)     & (95.6)     &      &      \\
 TDLSA & 1.00       & 1.00       & 1.00       & 1.00       & 1.00       & 1.00       & 1.00       & 1.00       &      &      \\
       & (95.0)     & (94.8)     & (96.8)     & (94.4)     & (97.0)     & (95.6)     & (95.4)     & (95.6)     &      &      \\
 SDLSA & 1.01       & -          & -          & 1.04       & -          & -          & 1.03       & -          & 3.00 & 1.00 \\
\bottomrule
\end{tabular}
\end{center}
\end{table}

\begin{table}
  \caption{Simulation results (with 100 replications) for logistic regression model by interleaving the bootstrap methods.  The REE of one-shot (OS),
  DLSA, and bootstrap estimators are reported.  }\label{bootstrap}
\begin{center}
\begin{tabular}{ccccccccccc}
\toprule
 $N$ & $K$ & Est.      & $\theta_1$ & $\theta_2$ & $\theta_3$ & $\theta_4$ & $\theta_5$ & $\theta_6$ & $\theta_7$ & $\theta_8$ \\
\hline
&&&\multicolumn{8}{c}{\sc Setting 1: i.i.d Covariates }                                                                          \\
5    & 8   & OS        & 0.70       & 0.97       & 0.97       & 0.78       & 0.94       & 0.97       & 0.73       & 0.96       \\
     &     & DLSA      & 0.90       & 1.03       & 1.02       & 0.90       & 1.03       & 1.02       & 0.95       & 1.02       \\
     &     & Bootstrap & 0.68       & 1.02       & 1.04       & 0.71       & 1.01       & 1.04       & 0.76       & 1.01       \\
20   & 10  & OS        & 0.90       & 0.99       & 0.99       & 0.91       & 0.98       & 1.00       & 0.92       & 0.99       \\
     &     & DLSA      & 0.89       & 1.00       & 1.01       & 0.97       & 1.00       & 1.00       & 0.92       & 1.01       \\
     &     & Bootstrap & 0.86       & 0.97       & 1.00       & 0.95       & 1.00       & 0.99       & 0.90       & 0.99       \\
\hline
&&&\multicolumn{8}{c}{\sc Setting 2: Heterogenous Covariates }                                                                   \\
5    & 8   & OS        & 0.54       & 0.85       & 0.82       & 0.63       & 0.78       & 0.87       & 0.62       & 0.77       \\
     &     & DLSA      & 0.77       & 1.02       & 1.02       & 0.86       & 1.04       & 1.04       & 0.84       & 1.04       \\
     &     & Bootstrap & 0.52       & 0.97       & 1.00       & 0.65       & 0.99       & 1.01       & 0.61       & 1.05       \\
20   & 10  & OS        & 0.68       & 0.88       & 0.99       & 0.79       & 0.91       & 0.91       & 0.75       & 0.89       \\
     &     & DLSA      & 0.90       & 1.02       & 1.01       & 1.01       & 1.00       & 1.01       & 0.92       & 1.02       \\
     &     & Bootstrap & 0.83       & 1.04       & 0.99       & 0.98       & 0.98       & 0.99       & 0.86       & 1.00       \\
\bottomrule
\end{tabular}
\end{center}
\end{table}

\begin{table}
  \caption{Simulation results for screening result with 100 replications.  The CRs of 4 screening methods are reported.  }\label{screening_cp}
\begin{center}
\begin{tabular}{rcccccccccc}
\toprule
$N$   & $K$ & SIS  & SIRS & KC   & DC && SIS  & SIRS & KC   & DC                      \\

      &   & \multicolumn{4}{c}{\sc Linear Regression} &      & \multicolumn{4}{c}{\sc Logistic Regression} \\
  \cline{3-6}  \cline{8-11}
      &   & \multicolumn{9}{c}{\sc Setting 1: i.i.d Covariates }                                           \\
2000  & 5 & 1.00                                      & 0.71 & 0.87 & 0.91 &  & 0.99 & 0.48 & 0.73 & 0.89  \\
10000 & 8 & 1.00                                      & 0.92 & 1.00 & 0.99 &  & 1.00 & 0.81 & 0.73 & 0.99  \\

      && \multicolumn{9}{c}{\sc Setting 2: Heterogenous Covariates } \\
2000  & 5   & 0.98 & 0.73 & 0.79 & 0.84 && 0.95 & 0.55 & 0.73 & 0.85                     \\
10000 & 8   & 1.00 & 0.86 & 0.91 & 0.95 && 1.00 & 0.77 & 0.73 & 0.94                     \\
\bottomrule
\end{tabular}
\end{center}
\end{table}

\begin{table}
  \caption{Simulation results for linear and logistic regression models after screening  procedure (using SIS) with 100 replications.
  The numerical
    performances are evaluated for different sample sizes $N$ ($\times 10^3$) and numbers
    of workers $K$.
  The CR$_{screen}$, REE (of five modelling methods for all covariates),
  CR$_{select}$, CM and are reported.  }\label{linear_after_scr}
\begin{center}
\begin{tabular}{cccccccccc}
\toprule
 $N$ & $K$ & CR$_{screen}$ & OS   & CSL  & DLSA & TDLSA & SDLSA & CR$_{select}$ & CM   \\
\midrule
     && \multicolumn{8}{c}{\sc Linear Regression}                                 \\
&& \multicolumn{8}{c}{\sc Setting 1: i.i.d Covariates }                                 \\
 2   & 5   & 1.00          & 0.97 & 0.92 & 1.00 & 1.00  & 1.02  & 0.97          & 0.97 \\
 10  & 8   & 1.00          & 0.98 & 0.92 & 1.00 & 1.00  & 1.02  & 1.00          & 1.00 \\
&& \multicolumn{8}{c}{\sc Setting 2: Heterogenous Covariates }                          \\
 2   & 5   & 0.98          & 1.00 & 1.00 & 1.00 & 1.00  & 1.00  & 0.74          & 0.74 \\
 10  & 8   & 1.00          & 0.98 & 0.87 & 1.00 & 1.00  & 1.01  & 1.00          & 1.00 \\
\cline{3-10}
     && \multicolumn{8}{c}{\sc Logistic Regression}                                 \\

&& \multicolumn{8}{c}{\sc Setting 1: i.i.d Covariates }                                 \\
 2   & 5   & 0.99          & 0.18 & 0.03 & 0.45 & 1.09  & 0.96  & 0.87          & 0.82 \\
 10  & 8   & 1.00          & 0.37 & 0.30 & 0.58 & 1.07  & 1.08  & 1.00          & 0.98 \\
&& \multicolumn{8}{c}{\sc Setting 2: Heterogenous Covariates }                          \\
 2   & 5   & 0.95          & 0.55 & 0.05 & 0.71 & 0.95  & 0.91  & 0.42          & 0.39 \\
 10  & 8   & 1.00          & 0.44 & 0.32 & 0.58 & 1.02  & 0.98  & 0.99          & 0.98 \\
 \bottomrule
\end{tabular}
\end{center}
\end{table}

\begin{table}
  \caption{Simulation results for comparison of the shrinkage estimation with the hypothesis testing approach with 500 replications.  The coverage rate (CR) and percentage of identifying the true model (CM) are presented.}\label{tab_shrinkage}
\begin{center}
\begin{tabular}{ccccccccccccc}
\toprule
      &     & \multicolumn{5}{c}{\sc Setting 1: i.i.d Case } &      & \multicolumn{5}{c}{\sc Setting 2: Heterogenous Case}                                                      \\
\cline{3-7} \cline{9-13}
      &     & \multicolumn{2}{c}{\sc Testing}                &      & \multicolumn{2}{c}{\sc SDLSA} &      & \multicolumn{2}{c}{\sc Testing} &  & \multicolumn{2}{c}{\sc SDLSA} \\
\cline{3-4} \cline{6-7} \cline{9-10} \cline{12-13}
  $N$ & $K$ & CR                                             & CM   &                               & CR   & CM                              &  & CR   & CM   &  & CR   & CM    \\
\midrule
  20  & 5   & 1.00                                           & 0.47 &                               & 1.00 & 0.94                            &  & 1.00 & 0.47 &  & 1.00 & 0.90  \\
100   & 10  & 1.00                                           & 0.49 &                               & 1.00 & 0.99                            &  & 1.00 & 0.46 &  & 1.00 & 0.99  \\
\bottomrule
\end{tabular}
\end{center}
\end{table}


\end{document}